\documentclass[aps,prd,showpacs,amssymb,floatfix,nofootinbib,preprintnumbers]{revtex4}
\newcommand{\be}{\begin{equation}}
\newcommand{\ee}{\end{equation}}
\newcommand{\beq}{\begin{eqnarray}}
\newcommand{\eeq}{\end{eqnarray}}
\usepackage{bm}
\usepackage{graphicx}%
\usepackage{epsfig}
\usepackage{amsmath}
\usepackage{amsfonts,amssymb}
\usepackage{hhline}
\usepackage{multirow}

\newcommand{\dslash}{\not{\hbox{\kern-2pt $\partial$}}}
\newcommand{\tr}{\mbox{tr}}
\newcommand{\Tr}{\mbox{Tr}}

\begin{document}
\preprint{CP3-Origins-2010-13, WUB/10-07}
\title{The infrared dynamics of Minimal Walking Technicolor}
\author{Luigi Del Debbio}
\email[]{luigi.del.debbio@ed.ac.uk}
\affiliation{SUPA, School of Physics and Astronomy, University of Edinburgh,
	Edinburgh EH9 3JZ, Scotland}
\author{Biagio Lucini}
\email[]{b.lucini@swansea.ac.uk}
\author{Agostino Patella}
\email[]{a.patella@swansea.ac.uk}
\affiliation{School of Physical Sciences, Swansea University,
Singleton Park, Swansea SA2 8PP, UK}
\author{Claudio Pica}
\email[]{pica@cp3.sdu.dk}
\affiliation{CP$^3$-Origins, University of Southern Denmark
  Odense, 5230 M, Denmark}
\author{Antonio Rago}
\email[]{rago@physik.uni-wuppertal.de}
\affiliation{Department of Physics, Bergische Universit\"at Wuppertal, Gaussstr. 20, D-42119 Wuppertal, Germany}
\begin{abstract}
We study the gauge sector of {\em Minimal Walking Technicolor}, which
is an SU(2) gauge theory with $n_f = 2$ flavors of Wilson fermions in
the adjoint representation. Numerical simulations are performed on
lattices $N_t \times N_s^3$, with $N_s$ ranging from 8 to 16 and $N_t
= 2 N_s$, at fixed $\beta = 2.25$, and varying the fermion bare mass
$m_0$, so that our numerical results cover the full range of fermion
masses from the quenched region to the chiral limit. We present
results for the string tension and the glueball spectrum. A comparison
of mesonic and gluonic observables leads to the conclusion that the
infrared dynamics is given by an SU(2) pure Yang-Mills theory with a
typical energy scale for the spectrum sliding to zero with the fermion
mass. The typical mesonic mass scale is proportional to, and much
larger than this gluonic scale. Our findings are compatible with a
scenario in which the massless theory is conformal in the infrared. An
analysis of the scaling of the string tension with the fermion mass
towards the massless limit allows us to extract the chiral condensate
anomalous dimension $\gamma_*$, which is found to be $\gamma_* = 0.22
\pm 0.06$.
\end{abstract}
\pacs{11.15.Ha, 12.60.Nz, 12.39.Mk, 12.39.Pn}
\maketitle

\section{Introduction}
\label{sect:1}
A possible mechanism of Electroweak Symmetry Breaking is provided by
strongly-interacting dynamics beyond the Standard Model
(BSM)~\cite{Weinberg:1975gm,Susskind:1978ms}. In this picture, a new
strongly-coupled gauge interaction acting at energy scales of the
order of 1 TeV is conjectured. This interaction embeds the Standard
Model gauge group $SU(2)_L \otimes U(1)_Y$ and contains fermionic
degrees of freedom (different from the Standard Model fermions) that
do not interact directly with the latter. The chiral symmetry of the
BSM interaction breaks spontaneously at the scale of 1 TeV. This
breaking provides mass to the $Z$ and $W^{\pm}$ bosons of the Standard
Model. Historically, this framework is known as Technicolor; the new
bosons are referred to as technibosons and the new fermions as
technifermions. In order to give mass to the Standard Model fermions,
another new gauge interaction acting at higher energy scales ({\em
  Extended Technicolor}) is introduced.
 
In the original proposals the strongly interacting BSM dynamics was
obtained by rescaling QCD. The ratio between the scale of the
technicolor model and the QCD scale can then be used to determine the
mass spectrum of the BSM theory. However, this scenario proves to be
inadequate to provide a mechanism of mass generation for fermions
without running into problems with flavor changing neutral currents. A
more refined framework that could avoid those problems is Walking
Technicolor~\cite{Holdom:1984sk,Yamawaki:1985zg,Appelquist:1986an}. 

Walking theories are realized as deformations of theories with an
infrared (IR) fixed point, i.e. a point in which the $\beta$ functions
for the couplings of the theory
vanish~\cite{Caswell:1974gg,Banks:1981nn}.
The role played in this scenario by the fermion representation has
been emphasized in Refs.~\cite{Sannino:2004qp,Dietrich:2006cm}. In
particular, for theories in the two-index symmetric and adjoint
representations, an IR fixed point can be reached at smaller values of
the number of fermion flavors $n_f$ than for theories involving
fundamental fermions. 

Like any other BSM framework, Technicolor has to confront the
stringent experimental bounds for new physics summarized in the $S$
and $T$ parameters~\cite{Peskin:1991sw}. Recent reviews of the
phenomenological aspects of Technicolor theories can be found in
Refs.~\cite{Hill:2002ap,Lane:2002wv,Sannino:2008ha,Sannino:2009za,Piai:2010ma}.
It is currently an open question whether a theoretically consistent
framework for Electroweak Symmetry Breaking can be drawn from those
ideas. In particular, one would like to explore from first principles
whether SU($N$) gauge theories with $n_f$ fermion flavors in the
fundamental or in a two-index representation can provide a viable
walking scenario for some values of $N$ and $n_f$. Such theories would
be natural candidates as models of strongly interacting BSM dynamics,
which eventually will be tested at the LHC. 

Ultimately, the issue of determining the features of a SU($N$) gauge
theory coupled with $n_f$ fermion flavors transforming according to
some representation ${\cal R}$ of the gauge group is of a
nonperturbative nature, and as such it can be studied in the framework
of lattice gauge theories (see e.g.~\cite{Nunez:2008wi} for a
complementary approach based on AdS/CFT techniques). Following the
work of Ref.~\cite{Catterall:2007yx}, other lattice studies have
focused on theories with fermions in two-index representations
conjectured to be relevant as models of strongly interacting BSM
dynamics: preliminary results have appeared for SU(2) with two
fermions in the adjoint
representation~\cite{DelDebbio:2008zf,Catterall:2008qk,Hietanen:2008mr,Pica:2009hc,Catterall:2009sb},
and SU(3) with two fermions in the symmetric
representation~\cite{DeGrand:2008kx}. These studies found a mass
spectrum characterized by the degeneracy of the pseudoscalar and the
vector meson in a wide range of fermion masses. Complementary
investigations of the running of the
coupling~\cite{Shamir:2008pb,Svetitsky:2009pz,Hietanen:2009az,Hietanen:2009zz,Bursa:2009we,Bursa:2009tj}, and of the
exponents that govern the scaling towards the massless
limit~\cite{DeGrand:2009hu,DeGrand:2009mt,DeGrand:2009et,Lucini:2009an} found
preliminary indication for the existence of an infrared fixed
point. Note that while in SU(2) with adjoint fermions there seems to
be consensus that the theory has an infrared conformal fixed point,
for SU(3) with sextet fermions the study of Ref.~\cite{Fodor:2009ar}
supports a QCD-like scenario, while Ref.~\cite{Kogut:2010cz,Sinclair:2009ec} favors a
walking scenario. A closely related line of research is the lattice
determination of the critical number of flavors for the onset of the
conformal window in the SU(3) gauge theory with fermions in the fundamental
representation~\cite{Appelquist:2007hu,Deuzeman:2008sc,Appelquist:2009ty,
Bilgici:2009kh,Deuzeman:2009mh,Fodor:2009wk,Fodor:2009rb,Hasenfratz:2009ea,
Hasenfratz:2009kz,Jin:2009mc,Appelquist:2009ka,Hasenfratz:2010fi}.
Ref.~\cite{DelDebbio:2008wb} provides a perturbative
determination of renormalization constants relating lattice and continuum
observables in SU($N$) gauge theories with fermions in two-index
representations. A recent account of the activity in the field is provided in
Refs.~\cite{Fleming:2008gy,Lucini:2009an,Pallante:2009hu}. For current
numerical studies of conformal gauge theories, it proves to be helpful
to have analytical estimates of the extent of the conformal window;
recent works on this subject are reported in
Refs.~\cite{Poppitz:2009uq,Armoni:2009jn,Sannino:2009me,Poppitz:2009tw}.

Numerical simulations of the spectrum of candidate theories of
Electroweak Symmetry Breaking beyond the Standard Model have focused
almost exclusively on the meson spectrum (and in particular, on the
states that in QCD are the lowest-lying particles of the meson
isovector spectrum, namely the pseudoscalar and the vector mesons). Recently,
investigating the case of a SU(2) gauge theory with two fermion
flavors in the adjoint representation (which is commonly referred to
as {\em Minimal Walking Technicolor}), we have pointed out in
Ref.~\cite{DelDebbio:2009fd} that a clean signature of conformality in
the chiral limit can be obtained by comparing mesonic and gluonic
observables. In particular, a conformal gauge theory broken with a
small fermion mass term displays the phenomenon of \textit{hyperscaling}
and \textit{locking},
i.e. all the ratios of spectral quantities are independent of the
fermion mass if the latter is sufficiently small. This paper has the
twofold motivation of discussing more extensively the general
expectations for the spectrum when a small mass term breaks explicitly
conformal invariance, and of presenting the details of our analysis of
the data in the gluonic sector leading to the conclusions of
Ref.~\cite{DelDebbio:2009fd} about the likely existence of an infrared
fixed point,
having increased the statistics at some values of the lattice
parameters. This work complements the investigation reported
in Ref.~\cite{noi}, where our results for mesonic observables were
discussed. We shall use the evidence found for the theory to have an
infrared fixed point to perform a scaling analysis of our observables
as a function of the fermion mass and provide an estimate for the
anomalous dimension of the condensate, whose value has relevant
phenomenological implications. 

As in Refs.~\cite{DelDebbio:2009fd,noi}, the study reported here is at
fixed lattice spacing. One key issue that should be carefully
discussed is whether our results are relevant for the continuum
physics. In general, the program of extracting the values of
observables in the continuum from lattice simulations of BSM models is
still at an early stage; in practice, numerical results are obtained
for volumes and lattice couplings that are argued to be a good
approximations of the continuum system. Assessing the reliability of
lattice simulations for continuum physics requires then a detailed
knowledge of the phase structure of the lattice theory. More in
detail, it is easy to prove analytically that deep in the strong
coupling phase a SU($N$) lattice pure gauge theory is always
confined. This feature survives when fermions with sufficiently high
mass are added to the action. The lattice strong coupling, also known
as the bulk phase, is separated from the continuum phase by either a
phase transition or a smooth crossover (for a study of the strong
coupling regime with fermions in the two-index representation,
see~\cite{Moraitis:2009xt}). Note that the latter phase may
or may not be confining. In order to obtain a reliable continuum
extrapolation, only points for which the system is in the same phase
as the continuum theory must be considered. Hence, one of the
preliminary tasks of lattice simulations is to identify the exact
extent of the bulk phase. This program has been carried out in
Refs.~\cite{DelDebbio:2008zf,Catterall:2008qk,Hietanen:2008mr}, which
have shown that the bulk phase roughly corresponds to lattice
couplings $\beta \le 2.0$. However, staying clear from the bulk phase
could not be a sufficient condition for getting relevant results for
the real-world physics: another aspect that needs to be considered is
the physical size of the volume, which should be such that analytical
predictions for finite size corrections could be reliably used to
extract information from the data. Although at first sight this issue
could seem more under control than the one related to lattice
artefacts, this is a prejudice modeled after our understanding of
QCD. If the physics of our system is conformal, QCD does not provide a
reliable guidance for analyzing the numerical results. In fact, recent
analytical and numerical arguments have shown that, even in the
continuum, the theory in a finite box is characterized by a
non-trivial phase structure in terms of the lattice volume and of the
fermion mass~\cite{Myers:2009df,Cossu:2009sq,Machtey:2009wu}. The finite volume
phases are characterized by different behaviors of the order
parameter related to confinement, the trace of the Polyakov loop. In
order to check for possible lattice artefacts, we measured the
Polyakov loop wrapping around the spatial directions. We find that the
bare fermion mass can significantly affect the phase structure at
fixed lattice size. In particular, at fixed volume, when lowering the
bare fermion mass the system goes from the infinite-volume confined
phase (as shown by the presence of a non-zero string tension) to a
spatially deconfined regime. We find that gluonic observables are
strongly affected by this change of regime, while mesonic quantities
behave smoothly across it. This could be an effect of the separation
at finite mass of the confinement and the chiral symmetry breaking
scales in theories with adjoint fermions (observed
in Refs.~\cite{Kogut:1985pp,Karsch:1998qj}), which would be an expected
feature for a candidate model of Technicolor that has a large distance
dynamics different from the QCD one.

This work is organized as follows. Sect.~\ref{sect:rg} discusses the
hyperscaling and locking phenomena, which arises when an infrared conformal gauge
theory is deformed with a small mass term. In Sect.~\ref{sect:lattice}
we define the discretized theory and set the notations. Results for
the phase structure, the string tension extracted via Polyakov loop
correlators, the string tension extracted from expectation values of
Wilson loops and glueball masses are reported in
Sect.~\ref{sect:lattice:center}-\ref{sect:lattice:glueballs}. A
comparison between the dynamical and the quenched simulations
performed keeping the string tension and the pseudoscalar mass fixed
at the values dictated by the dynamical theory is then provided, and
from this comparison hyperscaling and locking are shown to take place in the model
studied in this work
(Sect.~\ref{sect:locking}). Sect.~\ref{sect:scaling} illustrates our
scaling analysis aimed to determine the chiral condensate anomalous
dimension $\gamma_*$. Finally, Sec.~\ref{sect:conclusions} reports our
conclusions and possible future directions of our work.
\section{Mass-deformed infrared-conformal gauge theories}
\label{sect:rg}
For technical reasons which depend on the specific fermion
discretization, lattice simulations can only be performed with a
non-vanishing mass term for the fermions. In particular Wilson
fermions break chiral symmetry explicitly even for vanishing bare
mass, so that the massless limit is only obtained by fine-tuning the
parameters in the Lagrangian. Moreover, it is impossible to simulate
at arbitrarily small masses if the lattice spacing and the volume are
kept constant, since small eigenvalues of the Dirac operator are
generated, the simulation algorithm becomes unstable and unphysical
phases can appear. The extrapolation from a region of small enough
masses (but still in a safe region of parameters) to the chiral limit
can be performed only under the guidance of an analytical picture.

For QCD-like theories, chiral perturbation theory (in the infinite
volume, in the epsilon and delta regimes) allows to extrapolate
physical quantities from a region of small enough masses to the chiral
limit.

The natural question is: what should we expect if we deform an
IR-conformal theory with a small mass term, and how do we recover the
chiral limit? Hence, before illustrating the details of our
simulations, we set the frame for the picture in the latter case.

\subsection{Renormalization Group analysis}
\label{subsect:rg}
A gauge theory with massless fermions (in the continuum) depends on a
single parameter, the running coupling $g(\mu_0)$ at some reference
scale $\mu_0$, or alternatively the RG-invariant parameter
$\Lambda$. This is valid for both confining and IR-conformal
theories. In confining theories the particle masses (except the
Goldstone bosons) are proportional to the parameter $\Lambda$ in the
chiral limit. In the case of IR-conformal theories, where the spectrum
is made of unparticles, $\Lambda$ is not associated to particle
masses, but sets the energy scale at which the cross-over between the
asymptotically free and conformal regimes occurs. An explicit
definition of $\Lambda$ is not relevant for our discussion and will
then be omitted.

When the IR-conformal theory is deformed by a mass term for the
fermions, a particle spectrum with a mass gap is expected to be
generated. The theory depends now on one more parameter, the running
mass $m(\mu_0)$ at the reference scale $\mu_0$; alternatively, an
RG-invariant parameter $M$ can be suitably defined. Close enough to
the chiral limit (in the \textit{scaling region}), the particle masses
are expected to be independent of $\Lambda$. We will see that under a
regularity hypothesis, those masses are expected to be proportional to
$M$. This result is standard in the statistical-mechanics analysis of
second-order phase transitions, but it will be presented here using
the language of quantum field theory.

The running of the renormalized mass is computed by solving the RG
equation (in a mass-independent renormalization scheme):
\begin{equation}
\mu \frac{d m}{d \mu}(\mu) = - \gamma(g(\mu)) m(\mu) \ ,
\end{equation}
which yields:
\begin{equation} 
  \label{eq:runningmass} 
  m(\mu) = m(\mu_0) \exp \left\{
    - \int_{g(\mu_0)}^{g(\mu)} \frac{\gamma(z)}{\beta(z)} dz \right\}
  \equiv Z_m(\mu,\mu_0,\Lambda) m(\mu_0) \ .
\end{equation}
As we are going to show, the function $Z_m(\mu,\mu_0,\Lambda)$ can be
rewritten in a more convenient form. The theory we are interested in
is asymptotically free in the UV. The $\beta$ and $\gamma$
functions close to the UV fixed point are:
\begin{flalign}
g \to 0 \ : \ & \beta(g) \simeq - \beta_0 g^3 \ , \\
& \gamma(g) \simeq \gamma_0 g^2 \ ,
\end{flalign}
where the lowest order coefficients come from a one-loop computation
($T_R$ and $C_2(R)$ are the generator normalization and the Casimir of
the fermionic representation):
\begin{gather}
  \beta_0 = \frac{1}{(4 \pi)^2} \left( \frac{11}{3} N - \frac{4}{3} T_R n_F \right) \ , \\
  \gamma_0 = \frac{6C_2(R)}{(4 \pi)^2} \ .
\end{gather}
>From now on, we will be interested only in the IR-conformal scenario.
Close to the IR fixed point we assume a regular behavior for the RG
functions:
\begin{flalign}
g \to g_* \ : \ & \beta(g) \simeq \beta_* ( g - g_* ) \ , \\
& \gamma(g) \simeq \gamma_* \ ,
\end{flalign}
where $\beta_*$ and $\gamma_*$, which are scheme-independent
quantities, are in general not accessible by a perturbative
expansion.

Integrating the RG equation $\mu dg/d\mu = \beta(g)$ close to the
fixed points, the asymptotic running-coupling behavior is derived:
\begin{flalign}
  \mu \to \infty \ : \ & g(\mu) \simeq \frac{1}{2 \beta_0 \log ( \mu / \Lambda )} \ , \\
  \mu \to 0 \ : \ & g(\mu) \simeq g_* - A_g \left( \frac{\mu}{\Lambda}
  \right)^{\beta_*} \ .
\end{flalign}

We separate now the singular behaviors close to the fixed points in
the multiplicative renormalization function of the mass:
\begin{flalign}
Z_m(\mu,\mu_0,\Lambda)
= & \exp \left\{ - \int_{g(\mu_0)}^{g(\mu)} \left( \frac{\gamma(z)}{\beta(z)} - \frac{\gamma_*}{\beta_*(z-g_*)} + \frac{\gamma_0}{\beta_0 z} \right) dz \right\} \times \nonumber \\
& \times \exp \left\{ - \int_{g(\mu_0)}^{g(\mu)} \frac{\gamma_*}{\beta_*(z-g_*)} dz \right\}
\exp \left\{ \int_{g(\mu_0)}^{g(\mu)} \frac{\gamma_0}{\beta_0 z} dz \right\} = \nonumber \\
= & \frac{ \tilde{Z}_m(\mu / \Lambda) }{ \tilde{Z}_m(\mu_0 / \Lambda) } \ , \label{eq:zeta_factorization}
\end{flalign}
where the function
\begin{equation}
  \tilde{Z}_m(\mu / \Lambda) =
  [ g_*-g(\mu) ]^{-\frac{\gamma_*}{\beta_*}}
  g(\mu)^{\frac{\gamma_0}{\beta_0}}
  \exp \left\{ \int_{g(\mu)}^{g_*} \left( \frac{\gamma(z)}{\beta(z)} - \frac{\gamma_*}{\beta_*(z-g_*)} + \frac{\gamma_0}{\beta_0 z} \right) dz \right\}
\end{equation}
is defined in such a way that the integral in the exponential is
finite both for $\mu \to 0$ and $\mu \to \infty$.

An RG-invariant fermionic mass $M$ can be defined by means of the
condition $m(M) = M$. Plugging Eq.~\eqref{eq:zeta_factorization} in
Eq.~\eqref{eq:runningmass}, and choosing $\mu_0=M$ we get the
relationship:
\begin{equation} \label{eq:runningmass_M}
\tilde{Z}_m(\mu/\Lambda)^{-1} m(\mu) = 
\tilde{Z}_m(M/\Lambda)^{-1} M \, .
\end{equation}

If the RG-invariant mass $M$ is much larger than $\Lambda$, the
following asymptotic behavior can be easily shown to hold by using the
previous relationship:
\begin{equation}
  m(\mu) = A_\infty \tilde{Z}_m(\mu/\Lambda) M \left[\log\frac{M}{\Lambda} 
  \right]^{\frac{\gamma_0}{\beta_0}} \, .
\end{equation}
At fixed energy scale $\mu$, the running mass diverges as $M$ goes to
infinity. The fermions decouple and the theory is effectively
described by a pure Yang-Mills with a scale $\Lambda_\mathrm{YM} \simeq
\Lambda$. At leading order in $\Lambda/M$, the parameter $M$
coincides with the quark pole mass. In fact, if $S(p)$ is the quark
propagator in a fixed gauge, the perturbative expansion yields:
\begin{equation}
  S(p)^{-1} = \left[ 1 + \sum_{n=1}^{\infty} A_n \left( \frac{-p^2}{\mu^2}, \frac{m(\mu)}{\mu} \right) g^{2n}(\mu) \right] \left[ \not{p} - m(\mu) - \sum_{n=1}^{\infty} B_n \left( \frac{-p^2}{\mu^2}, \frac{m(\mu)}{\mu} \right) g^{2n}(\mu) \right] \ ,
\end{equation}
the pole mass $\bar{m}$ is defined in such a way that the quark
propagator has a pole for $-p^2 = \bar{m}^2$. The pole mass is RG
invariant, therefore it can be computed for an arbitrary value of
$\mu$. It is convenient to choose $\mu=M$:
\begin{equation}
  \bar{m} = M + \sum_{n=1}^{\infty} B_n \left( \frac{\bar{m}^2}{M^2}, 1 \right) g^{2n}(M) \ .
\end{equation}
At large masses $M \gg \Lambda$, the terms in the sum are suppressed
since the running coupling goes to zero, and $\bar{m} \simeq M$. In
this regime, the meson masses are just twice the quark pole mass,
while the glueball masses are the same as in the pure Yang-Mills
theory:
\begin{gather}
M_\mathrm{mes} = 2 M \ ; \\
M_\mathrm{glue} = B_\mathrm{glue} \Lambda \ .
\end{gather}

On the other hand, the chiral limit is reached for values of $M$ much smaller than
$\Lambda$. In this case, Eq.~\eqref{eq:runningmass_M} becomes:
\begin{equation} 
  \label{eq:runningmass_chiral}
  m(\mu) = A_0 \tilde{Z}(\mu/\Lambda) \Lambda^{-\gamma_*} M^{1+\gamma_*} \ ,
\end{equation}
producing the power law that is characteristic of the IR fixed point
deformed with a small fermionic mass.

Consider now a physical mass $M_X$ in a channel $X$ (it can be the
mass of a particle or other physical quantities like the square root
of the string tension). As every observable, this will be a function
of the renormalized coupling $g(\mu)$, the mass $m(\mu)$, and the
subtraction scale $\mu$. However a physical quantity must be RG
invariant:
\begin{equation} \label{eq:MX_mu}
M_X[\mu,g(\mu),m(\mu)] = M_X
\end{equation}
for every value of $\mu$. The RG equation for $M_X$ has a simple
solution in terms of the RG-invariant quantities $\Lambda$ and $M$:
\begin{equation} \label{eq:FX}
M_X = M \ F_X(M/\Lambda) \ ,
\end{equation}
where $F_X$ is a generic function of the ratio $M/\Lambda$. In
particular, if $F_X(x) = \alpha/x$, we get $M_X = \alpha \Lambda$
which is an RG-invariant quantity, but does not vanish in the chiral
limit.

The \textit{hyperscaling hypothesis}, which is assumed in the standard
discussion of second-order phase transitions (see
e.g. Ref.~\cite{Cardy:1996xt}), asserts the regularity of masses (or
correlation lengths in the language of statistical mechanics) with
respect to the irrelevant couplings. Consider Eq.~\eqref{eq:MX_mu} for
$\mu=M \ll \Lambda$:
\begin{flalign}
  M_X \simeq & M_X[M,g_*-A_g(M/\Lambda)^{\beta_*} ,M ] = \nonumber \\
  = & M_X[1, g_*-A_g(M/\Lambda)^{\beta_*} , 1] \ M \simeq \nonumber \\
  \simeq & M_X[1, g_* , 1] \ M  \equiv A_X M \ , \label{eq:hyperscaling}
\end{flalign}
where we used dimensional analysis for the second line, and regularity
with respect to $g$ in the last one. Under the hyperscaling
hypothesis, RG-invariant IR quantities depend only on $M$ (and not on
$\Lambda$) close enough to the chiral limit. The hyperscaling
hypothesis constraints the $F_X$ function defined in Eq.~\eqref{eq:FX}
to be regular in the chiral limit:
\begin{equation} \label{eq:FX_x0}
  \lim_{x \to 0} F_X(x) = A_X \ .
\end{equation}
Since a mass gap is expected to be generated at nonzero values of $M$, $A_X$ must be different
from zero.

Combining Eqs.~\eqref{eq:hyperscaling}
and~\eqref{eq:runningmass_chiral} we get the power law for physical
masses close to the chiral limit:
\begin{equation} 
  \label{eq:powerlaw1}
  M_X = A_X M = A_X [ A_0 \tilde{Z}(\mu/\Lambda)]^{-\frac{1}{1+\gamma_*}} \Lambda^{\frac{\gamma_*}{1+\gamma_*}} m(\mu)^{\frac{1}{1+\gamma_*}} \ .
\end{equation}
We remind that this expression is valid for every value of $\mu$ as
long as $M\ll\Lambda$. In particular, the independence of $M_X$ of
$\Lambda$ is manifest at values $\mu \ll \Lambda$:
\begin{equation} 
  \label{eq:powerlaw2}
  M_X = A_X \mu^{\frac{\gamma_*}{1+\gamma_*}} m(\mu)^{\frac{1}{1+\gamma_*}} \ .
\end{equation}
If we interpret the RG in the Wilsonian sense and choose $\mu=a^{-1}$
to be the cutoff, Eq.~\eqref{eq:powerlaw1} yields the power law
dependence of physical masses on the bare quark mass $a M_X \propto (a
m_0)^{\frac{1}{1+\gamma_*}}$.

\begin{figure}[ht]
\centering
\includegraphics*[width=.8\textwidth]{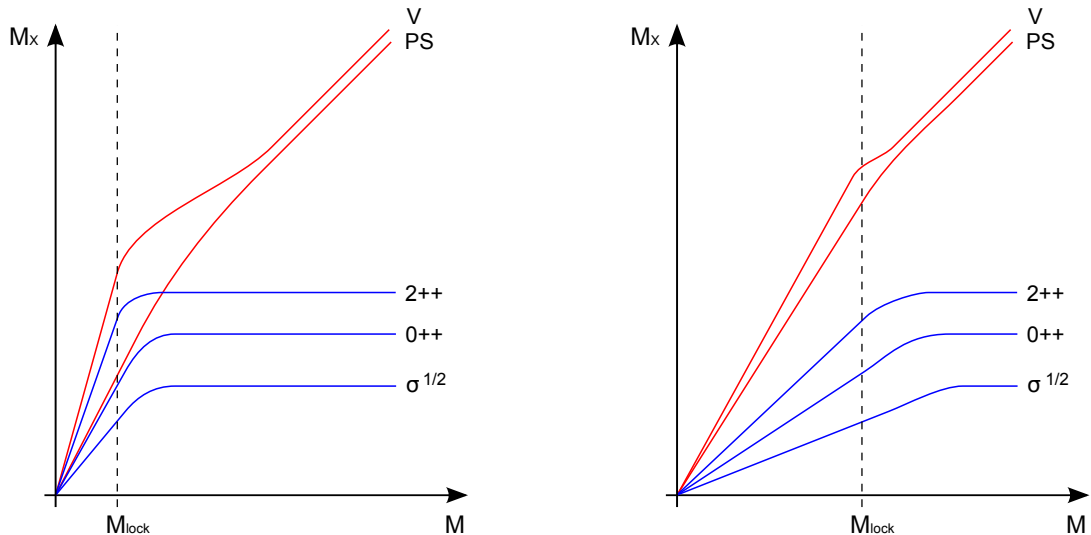}
\caption{Sketches of the spectrum of a mass-deformed IR-conformal
  theory (square root of the string tension, $0^{++}$ and $2^{++}$
  glueballs, pseudoscalar and vector isovector mesons). In the left
  plot, the locking sets up at an intermediate value of the fermion
  mass, where dynamical fermion effects account for the physics of the
  system, but the pseudoscalar is not much lighter than the other
  particles in the spectrum. In the right plot, the locking sets up at
  a high value of the fermion mass, where the heavy quark effective
  theory provides a good description of the relevant degrees of
  freedoms. This case is realized close to the Banks-Zacks point, but
  is possible in principle also if a strongly coupled IR fixed point
  is present.}
\label{fig:locking}
\end{figure}

\subsection{Scaling region and locking scale}
\label{subsect:locking}
Under the hyperscaling hypothesis, the function $F_X$ defined in
Eq.~\eqref{eq:FX} is expected to approach a nonzero value $A_X$ in the
chiral limit. We can define the scaling region for a given channel
$X$ as the range of $x=M/\Lambda$ around $x=0$, where the
function $F_X(x)$ deviates from its asymptotic behavior by a small
relative amount $\epsilon$:
\begin{equation}
\left| \frac{F_X(x) - A_X}{A_X} \right| < \epsilon \ .
\end{equation}
In the scaling region, the mass $M_X$ obeys the power
law~\eqref{eq:powerlaw1} as a function of the running mass up to
corrections of order $\epsilon$. The extension of the scaling region
will depend on the size of the discarded subleading
contributions to formula~\eqref{eq:powerlaw1} in the chosen channel.

Consider now the square root of the fundamental string tension
$M_\sigma=\sqrt{\sigma}$ (which is well defined for dynamical fermions
in the adjoint representation) and the lightest isovector meson (which
is always the pseudoscalar one), with mass $M_\mathrm{PS}$. A finite
value $x=\bar{x}$ exists, below which both these channels are in the
scaling region. This means that below the mass $M_\mathrm{lock} =
\bar{x} \Lambda$, the corrections to the hyperscaling behavior of
$M_{\sigma}$ and $M_\mathrm{PS}$ masses are relatively smaller than
$\epsilon$. Also the ratio $M_\mathrm{PS}/M_{\sigma}$ for every
fermionic mass below $M_\mathrm{lock}$ will be very similar to its
asymptotic value $A_\mathrm{PS}/A_{\sigma}$:
\begin{equation}
  \left| \frac{M_\mathrm{PS}}{M_{\sigma}} - \frac{A_\mathrm{PS}}{A_{\sigma}} \right| < O(\epsilon) \ .
\end{equation}

The dynamics is dramatically different below and above the mass
$M_\mathrm{lock}$. In the large-mass region, $M \gg \Lambda$, the
gluonic and mesonic masses are parametrically independent. All the
gluonic masses are proportional to $\Lambda$, while all the mesonic
masses are equal to $2M$:
\begin{gather}
  M_\mathrm{PS} = 2 M \ , \\
  M_{\sigma} = B_{\sigma} \Lambda \ .
\end{gather}
The ratio $M_\mathrm{PS}/M_\sigma$ goes to infinity in the large-mass
limit. For masses below $M_\mathrm{lock}$ the two masses
$M_\mathrm{PS}$ and $M_\sigma$ enter the scaling region, become both
independent of $\Lambda$ and proportional to $M$. The ratio
$M_\mathrm{PS}/M_\sigma$ is locked to its asymptotic value
$A_\mathrm{PS}/A_\sigma$. We will refer to $M_\mathrm{lock}$ as the
\textit{locking} mass.

The behavior of the masses in between the large-mass and scaling
regions and the actual value of $\bar{x}$ depend on the details of the
dynamics. However if the dynamics is such that the locking occurs at a
value $\bar{x} = M_\mathrm{lock}/\Lambda \gg 1$, then both $M_\sigma$
and $M_\mathrm{PS}$ at the locking scale are still approximately the
same as in the large-mass region:
\begin{gather}
  B_{\sigma} \simeq \frac{M_{\sigma}(M=M_\mathrm{lock})}{\Lambda} \simeq \frac{A_{\sigma} M_\mathrm{lock}}{\Lambda} = A_{\sigma} \bar{x} \ , \\
  A_\mathrm{PS} \simeq 2 \ ,
\end{gather}
and the ratio $M_\mathrm{PS}/M_\sigma$ is locked at a very large value:
\begin{equation}\label{eq:locking}
  \frac{A_\mathrm{PS}}{A_{\sigma}} \simeq \frac{2 \bar{x}}{B_{\sigma}} \gg 1\ .
\end{equation}

Mesons are much heavier than the square root of the string tension for
every value of $M$. Choosing an intermediate energy scale $E$ such
that $M_{\sigma} \ll E \ll M_\mathrm{PS}$, the effective theory
describing the gluonic degrees of freedom at energies below $E$ is a
pure Yang-Mills plus power-suppressed corrections coming from the
propagation of heavy quarks in the loops. In order to write the
effective Lagrangian in this regime, we need all the gauge-invariant
scalar operators of dimension 6 that are invariant under parity,
and charge conjugation. These can be written as linear combinations of
the following independent operators (a similar
analysis on the lattice was carried on in Ref.~\cite{Weisz:1982zw}):
\begin{eqnarray}
  \label{eq:dimsix1}
  S_1 &=& \sum_{\mu,\nu,\rho} \mathrm{tr } \left(
    J_{\mu\nu\rho} J^{\mu\nu\rho}\right) \, , \\
  \label{eq:dimsix2}
  S_2 &=& \sum_{\mu,\nu,\rho} \mathrm{tr } \left(
    J^\mu_{\phantom{\mu}\mu\rho} J_\nu^{\phantom{\nu}\nu\rho}\right) \, , \\
  \label{eq:dimsix3}
  S_3 &=& \sum_{\mu,\nu,\rho} \mathrm{tr } \left(
    J_{\mu\nu\rho} J^{\nu\mu\rho}\right) \, ,
\end{eqnarray}
where $J_{\mu\nu\rho}=\partial_\mu F_{\nu\rho} -i [A_\mu,F_{\nu\rho}]$.
Thus the effective Lagrangian can be written as:
\begin{equation}
  \mathcal{L}_\mathrm{eff} = -\frac{1}{2 g^2} \tr \left( F_{\mu\nu} F^{\mu\nu} \right) +
  \sum_{i=1,2,3}\frac{a_i}{M^2} S_i
  + O(M^{-4}) \ . \label{eq:efftheory}
\end{equation}

The scale $\Lambda_\mathrm{YM}$ of this low-energy pure Yang-Mills is in
general a function of $\Lambda$ and $M$ and can be computed by
matching the square root of the string tension of the low-energy
effective theory with the same quantity computed in the dynamical
theory:
\begin{equation}
  B_{\sigma} \Lambda_\mathrm{YM} \left[ 1 + O \left( \frac{\Lambda_\mathrm{YM}}{M} \right)^2 \right] = M_{\sigma} = M F_{\sigma}(M/\Lambda) \ ,
\end{equation}
which implies that trivially $\Lambda_\mathrm{YM} \simeq \Lambda$ for $M \gg
\Lambda$, while for $M < M_\mathrm{lock}$ then
\begin{equation} 
  \label{eq:sliding}
  \Lambda_\mathrm{YM} \simeq \frac{M_{\sigma}}{B_{\sigma}} \simeq \frac{A_{\sigma}M}{B_{\sigma}} \simeq \frac{M}{\bar{x}} \ .
\end{equation}
In the scaling region the scale $\Lambda_\mathrm{YM}$ of the low-energy pure
Yang-Mills \textit{slides} with the RG-invariant fermionic mass $M$.

A comment is mandatory at this point. \textit{At fixed value of the
  fermionic mass}, the low-lying spectrum of a mass-deformed
IR-conformal theory with $\bar{x} \gg 1$ can not be distinguished by
the low-lying spectrum of a confining theory with heavy quarks, since
they both are described by the same effective
Lagrangian~\eqref{eq:efftheory}. However in a genuine heavy-quark
phase the low-energy spectrum is almost independent of the mass $M$,
while the \textit{sliding} of the low-energy scale described in
Eq.~\eqref{eq:sliding} and (equivalently) the \textit{locking} of the
gluonic spectrum to the mass $M$ is ultimately a very clean signature
of IR-conformality.

Summarizing:
\begin{itemize}
\item We define the locking mass $M_\mathrm{lock}$ as the mass below which
  both the lowest isovector meson and the string tension are
  approximately in the chiral scaling region.
\item The value of $\bar{x}=M_\mathrm{lock}/\Lambda$ is determined by
  the detailed dynamics of the theory. If $\bar{x} \gg 1$ then the
  mesons are always much heavier than the square root of the string
  tension. The low-energy effective theory is a pure Yang-Mills plus
  small corrections, with a scale $\Lambda_\mathrm{YM}$ which depends
  on both $\Lambda$ and $M$. For $M > M_\mathrm{lock} \gg \Lambda$
  then the fermions completely decouple and $\Lambda_\mathrm{YM}
  \simeq \Lambda$, while for $M < M_\mathrm{lock}$ the only effect of
  the fermions in the dynamical theory is to make the low-energy scale
  slide with the fermionic mass $\Lambda_\mathrm{YM} \simeq M/\bar{x}$.
\item The case where $\bar{x} \gg 1$ is realized if the fixed point is
  perturbative~\cite{Miransky:1998dh,Miransky:2010kb}. In fact, in this case
\begin{equation}
  \bar{x} = \exp \left( \frac{1}{2 \beta_0^{YM} g_*^2} \right) \ .
\end{equation}
\end{itemize}

The described scenarios are illustrated in the sketches in
Fig.~\ref{fig:locking}.

\section{The lattice model}
\label{sect:lattice}
Consider a four-dimensional Euclidean torus $L_t \times L_s^3$, where
$L_t$ and $L_s$ are the lengths respectively of the temporal and
spatial directions. The space-time is discretized by introducing a
lattice with spacing $a$, and with $N_t=L_t/a$ and $N_s=L_s/a$ sites
respectively in the temporal and spatial directions. Lattice sites are
identified by four-coordinate dimensionful vectors
$x=(x_0,x_1,x_2,x_3)$. Therefore $x_0/a$ is an integer number running
from zero to $N_t-1$, and $x_i/a$ with $i=1,2,3$ are integer numbers
running from zero to $N_s-1$. In some cases it is useful to separate
the temporal coordinate from the spatial vector; we write $x=(t,\mathbf{r})$,
and $r$ is the modulus of ${\bf r}$. Lattice directions are indicated
with a Greek symbol and run from 0 to 3. The temporal direction is
chosen as the zero-th direction. We will use the same Greek symbol
both for the direction index and for the vector of length $a$ along
the axis direction (the meaning of the symbol will be always clear
from the context). The spatial directions are closed with periodic
boundary conditions (PBC) for all fields, while the boundary
conditions in the temporal direction are periodic for gauge fields and
antiperiodic (ABC) for fermion fields.

The action of a SU($N$) gauge theory with fermions can be decomposed as 
\begin{equation}
S = S_g + S_f \ , 
\end{equation}
where $S_g$ is the discretized Yang-Mills action and $S_f$ is the
fermionic contribution. Various choices for the lattice action are
possible, differing from each other by corrections that vanish in the
continuum limit. At finite lattice spacing different choices are
differently affected by lattice artefacts. In particular, as the
lattice spacing is increased, a transition to a phase not connected
with the continuum (the {\em bulk phase}) takes place. A careful
exploration of the phases of the system on a lattice as a function of
the lattice parameters is then mandatory.

For the gauge part, we use the Wilson action:
\begin{equation}
S_g = \beta \sum_{x,\mu<\nu} \left( 1 - \frac{1}{N} \textrm{Re} \, \tr {\cal P}_{\mu \nu}(x) \right) \ ,
\end{equation}
where ${\cal P}_{\mu \nu}(x)$ is the parallel
transport of the link variable $U(x,\mu) \in$ SU($N$) along the
elementary square of the lattice identified by the point $x$ and the
pair of directions $(\mu,\nu)$. $\beta$ is related to the
bare coupling $g_0^2$ by $\beta = 2 N/g_0^2$. The value of the
coupling determines the physical value of the ultraviolet cut-off, the
lattice spacing $a$. Note that independently of the fermion
representation, the link variables are in the fundamental
representation of SU($N$).

The fermion part of the action for a spinorial field $\psi(x)$ defined
on sites $x$ and transforming in the representation $R$ can be written
as
\begin{equation}
S_f =a^4  \sum_{k=1}^{n_f } \bar{\psi_k}(x) D_m \psi_k(x) \ ,
\end{equation}
where $D_m$ is the Dirac operator, in the Wilson discretization:
\begin{eqnarray}
D_m(x,y) = \left(\frac{4}{a} + m_0\right) \delta_{x,y} - \frac{1}{2a} \sum_\mu \left\{ \left(1-\gamma_\mu\right) U^R(x,\mu) \delta_{x,y-\mu} + \left(1+\gamma_\mu\right) U^R(y,\mu)^\dagger \delta_{x,y+\mu} \right\} \label{eq:Dirac}, 
\end{eqnarray}
where $U^R$ are the link variables in the representation $R$, and $m_0$ is the bare mass.

The functional integral is given by 
\begin{equation}
Z = \int \left( {\cal D} U \right) \left( {\cal D} \bar \psi \right) 
\left( {\cal D} \psi \right) e^{- S} = \int \left( {\cal D} U \right)
\left( \mbox{det} D_m \right)^{n_f} e^{- S_g}
\end{equation}
and the vacuum expectation value of an operator
$O(U,\psi,\bar{\psi})$ by
\begin{equation}
\langle O \rangle  = \frac{1}{Z} \int \left( {\cal D} U_{\mu} \right) \left( {\cal D} \bar \psi \right) \left( {\cal D} \psi \right) O e^{- S} \ ,
\end{equation}
where once again it is possible to integrate over the fermion fields
and obtain an expression that involves only an integral over the link
variables. For further details on the lattice formulation, we refer
to Refs.~\cite{DelDebbio:2008zf,noi}.

We performed numerical simulations for $SU(2)$ gauge theory with
$n_f=2$ Wilson fermions in the adjoint representation at fixed value
of $\beta=2.25$, different values of the bare mass, and different
lattices, the smallest one being a $16 \times 8^3$ lattice, and the
largest one being $32 \times 16^3$. We used the RHMC
algorithm~\cite{Clark:2006fx} as implemented in the HiRep code, which
is described and benchmarked in detail in
Ref.~\cite{DelDebbio:2008zf}. The full list of the parameters we used
in our simulations can be found in
Tables~\ref{tab:poly:8}-\ref{tab:poly:16}. In this work we are mainly
interested in gluonic observables. For every choice of the parameters
we compute:
\begin{itemize}
\item the traced Polyakov loops in every direction, in order to
  identify the regime of the theory, as described in
  Sect.~\ref{sect:lattice:center};
\item the string tension by means of correlators of spatial and
  temporal Polyakov loops, as described in
  Sect.~\ref{sect:lattice:string};
\item the static force and potential, as described in
  Sect.~\ref{sect:lattice:wilson};
\item the glueball masses, as described in
  Sect.~\ref{sect:lattice:glueballs}.
\end{itemize}
We follow the convention that lattice observables are dimensionful, with the same dimension of the corresponding continuum observable. Of course we can measure only dimensionless ratios. For instance given a mass $m$, only the dimensionless quantity $am$ can be extracted from lattice simulations. Determining $a(\beta)$, and ultimately the physical value $m$, requires to set the physical scale using an appropriate observable. We will not perform the step of reinstating physical units, but shall leave $a$ (which is fixed in our case) as a parameter. However, the reader must
bear in mind that knowing the value of $a$ in terms of the quantities
entering the dynamics in the continuum is important in order to
confidently assess the relevance of a lattice simulation for continuum
physics. For this investigation, following the detailed exploration of
the phase structure of the theory performed in Ref.~\cite{Hietanen:2008mr},
we argue that large discretization artefacs are ruled out and we
postpone to future studies a systematic investigation of these
effects.

\begin{table}[ht]
\centering
\begin{tabular}{ccc|cc|ccc}
\hline
\multirow{2}{*}{lattice} & \multirow{2}{*}{V} & \multirow{2}{*}{$-a m_0$} & \multicolumn{2}{c|}{spatial center} & \multicolumn{3}{c}{string tension from Polyakov correlators}\\
& & & $N_{conf}$ & realization & $a\sigma_t^{1/2}$ & $a\sigma_s^{1/2}$ & $\sigma_s = \sigma_t$\\
\hline
\hline
S0 & $16\times 8^3$ & -0.5  & 8000 & S   & --          & --         & --  \\
S1 & $16\times 8^3$ & -0.25 & 8000 & S   & 0.4085(57)  & 0.393(11)  & yes \\
S2 & $16\times 8^3$ & -0    & 8000 & S   & 0.3998(57)  & 0.388(11)  & yes \\
S3 & $16\times 8^3$ & 0.25  & 8000 & S   & 0.328(23)   & 0.358(12)  & yes \\
S4 & $16\times 8^3$ & 0.5   & 8000 & S   & 0.3576(46)  & 0.347(11)  & yes \\
S5 & $16\times 8^3$ & 0.75  & 8000 & S   & 0.282(13)   & 0.2784(75) & yes \\
S6 & $16\times 8^3$ & 0.90  & 8000 & S   & 0.227(11)   & 0.2452(67) & yes \\
A0 & $16\times 8^3$ & 0.95  & 8000 & S   & 0.1974(83)  & 0.2218(35) & no  \\
A1 & $16\times 8^3$ & 0.975 & 8000 & ?   & 0.2066(97)  & 0.2094(49) & yes \\
A2 & $16\times 8^3$ & 1     & 8000 & ?   & 0.1960(85)  & 0.2252(62) & no  \\
A3 & $16\times 8^3$ & 1.025 & 8000 & A   & 0.1689(44)  & 0.2109(46) & no  \\
A4 & $16\times 8^3$ & 1.05  & 8000 & A   & 0.1679(47)  & 0.2074(38) & no  \\
A5 & $16\times 8^3$ & 1.075 & 6400 & A   & 0.1629(27)  & 0.20680(94)& no  \\
A6 & $16\times 8^3$ & 1.1   & 6400 & A   & 0.1553(28)  & 0.20443(82)& no  \\
A7 & $16\times 8^3$ & 1.125 & --   & --  & 0.1462(26)  & 0.20423(87)& no  \\
A8 & $16\times 8^3$ & 1.15  & --   & --  & 0.1368(20)  & 0.20402(67)& no  \\
A9 & $16\times 8^3$ & 1.175 & --   & --  & 0.1310(19)  & 0.2067(11) & no  \\
\hline
\end{tabular}
\caption{Results for the Polyakov loop distribution and the string
  tension, for all the simulations on the $16\times 8^3$ lattice. We
  analyzed the Polyakov loop distribution by using the number of
  configurations quoted in the 4th column (for the three lowest masses
  we could not safely discard the thermalization). In the 5th column, 'S'
  indicates that the distribution has a single maximum in zero, 'A'
  indicates that a double peak or an asymmetric peak is visible, '?'
  indicates that we cannot clearly distinguish between the two
  cases. The temporal (6th column) and spatial (7th column) string
  tensions computed from correlators of Polyakov loops are quoted. In
  the 8th column, 'yes' indicates that the data for the temporal and
  spatial string tensions have an overlap at $1 \sigma$. }
\label{tab:poly:8}
\end{table}

\begin{table}[ht]
\centering
\begin{tabular}{ccc|cc|ccc}
\hline
\multirow{2}{*}{lattice} & \multirow{2}{*}{V} & \multirow{2}{*}{$-a m_0$} & \multicolumn{2}{c|}{spatial center} & \multicolumn{3}{c}{string tension from Polyakov correlators}\\
& & & $N_{conf}$ & realization & $a\sigma_t^{1/2}$ & $a\sigma_s^{1/2}$ & $\sigma_s = \sigma_t$\\
\hline
\hline
B0  & $24\times 12^3$ & 0.95  & 9501 & S   & 0.225(11)   & 0.2181(23)  & yes \\
B1  & $24\times 12^3$ & 1     & 7951 & S   & 0.1785(57)  & 0.1882(40)  & yes \\
B2  & $24\times 12^3$ & 1.05  & 5620 & ?   & 0.1491(73)  & 0.1597(24)  & yes \\
B3  & $24\times 12^3$ & 1.075 & 4987 & A   & 0.1398(63)  & 0.1461(20)  & yes \\
B4  & $24\times 12^3$ & 1.1   & 4194 & A   & 0.1205(56)  & 0.1427(21)  & no  \\
B5  & $24\times 12^3$ & 1.125 & 4001 & A   & 0.1130(54)  & 0.1338(5)   & no  \\
B6  & $24\times 12^3$ & 1.15  & 1500 & A   & 0.0829(26)  & 0.1337(9)   & no  \\
B7  & $24\times 12^3$ & 1.175 & 5001 & A   & 0.0918(36)  & 0.1333(6)   & no  \\
B8  & $24\times 12^3$ & 1.18  & 4490 & A   & 0.0944(36)  & 0.1333(11)  & no  \\
B9  & $24\times 12^3$ & 1.185 & 4335 & A   & 0.0834(29)  & 0.1425(47)  & no  \\
B10 & $24\times 12^3$ & 1.19  & 4336 & A   & 0.0851(34)  & 0.1380(37)  & no  \\
\hline
\end{tabular}
\caption{As Table~\ref{tab:poly:8}, for the simulations on the $24\times 12^3$ lattice.}
\label{tab:poly:12}
\end{table}

\begin{table}[ht]
\centering
\begin{tabular}{ccc|cc|ccc}
\hline
\multirow{2}{*}{lattice} & \multirow{2}{*}{V} & \multirow{2}{*}{$-a m_0$} & \multicolumn{2}{c|}{spatial center} & \multicolumn{3}{c}{string tension from Polyakov correlators}\\
& & & $N_{conf}$ & realization & $a\sigma_t^{1/2}$ & $a\sigma_s^{1/2}$ & $\sigma_s = \sigma_t$\\
\hline
\hline
C0 & $32\times 16^3$ & 1.15  & 6145 & ?   & 0.0790(34) & 0.1029(19)  & no  \\
C1 & $32\times 16^3$ & 1.175 & 1871 & A   & 0.0966(78) & 0.10057(65) & yes \\
C2 & $32\times 16^3$ & 1.18  & 1500 & A   & 0.0648(33) & 0.1086(13)  & no  \\
C3 & $32\times 16^3$ & 1.185 & 1419 & A   & 0.0612(36) & 0.09953(13) & no  \\
C4 & $32\times 16^3$ & 1.19  & 1609 & A   & 0.0703(40) & 0.1021(13)  & no  \\
\hline
\end{tabular}
\caption{As Table~\ref{tab:poly:8}, for the simulations on the $32\times 16^3$ lattice.}
\label{tab:poly:16}
\end{table}

\section{Center symmetries and Polyakov loops}
\label{sect:lattice:center}
For a SU($N$) gauge theory with fermions in the adjoint
representation, the action has a $\mathbb{Z}_N^4$ invariance,
where each of the $\mathbb{Z}_N$ factors is associated with one
direction of the system. An observable that is not invariant under the
$\mathbb{Z}_N$ factor of the symmetry group associated to the
direction $\rho$ is the Polyakov loop operator in the fundamental
representation wrapping around $\rho$:
\begin{equation}
  \bar{P}_{\rho} = {\sum_{x}}' \prod_{n=0}^{N_{\rho} - 1} U(x + n \rho,\rho) \ ,
\end{equation}
where the primed sum runs over all points of the lattice slice at
$x_\rho=0$. Consider the system in a Euclidean manifold $\mathbb{R}^3
\times S^1$, in which the direction $\rho$ is compactified on the
$S^1$ circle and the other three directions extend to infinity. At
given radius of the $S^1$, the $\mathbb{Z}_N$ symmetry associated to
$\rho$ might either be a symmetry of the system or could be
spontaneously broken. For simplicity, let us take $\rho$ to be the
temporal direction; the inverse radius of the $S^1$ has then the
natural interpretation of the temperature of the system. In this case,
if the system is symmetric, the trace of $\bar{P}_{\rho}$ is equal to
zero and the system is confined at the given temperature. Conversely,
a non-zero $\langle \tr \bar{P}_{\rho} \rangle$ signals the breaking
of the $\mathbb{Z}_N$ symmetry associated with the direction $\rho$
and the system is deconfined. Analogously, if $\rho$ is a spatial
dimension, we call the broken phase {\em spatial deconfinement}.

Even if the theory is conformal in the chiral limit, a small mass
drives it off the attraction basin of the IR fixed point,
generating a mass gap. As for the pure gauge system, at finite
temperature the theory is expected to undergo a deconfinement
transition as the temporal direction is reduced down to a critical
value. Due to the periodic boundary conditions for fermions in space,
the reduction of a single spatial direction is more subtle. Recent
numerical simulations~\cite{Cossu:2009sq} have shown that in a SU(3)
theory with two staggered Dirac adjoint fermions the center symmetry
(which is intact at infinite volume) is first broken and then restored
again as the radius of a compactified spatial direction is shrunk. The
symmetry restoring transition happening when the radius is shrunk from
values that set the system in the broken phase is expected from the
one-loop perturbative calculation of Ref.~\cite{kovtun:2007py}. The
critical radii for the symmetry breaking and symmetry restoring
transitions depend in general on the mass. Although the different
discretization choice and the different gauge group might affect the
phase structure, it is possible that a similar behavior characterizes
SU(2) with two adjoint Dirac flavors of Wilson fermions.

When the system is in a compact domain, non-trivial phases can be
observed only at large $N$. In this case, a rich phase structure can
exist, since more than one $\mathbb{Z}_N$ can be broken at the same
time. For the pure gauge theory at large $N$, as the hypercubic volume
is reduced from large values, a cascade of phase transitions takes
place~\cite{Narayanan:2007fb}. Each phase can be characterized by the
number of the $\mathbb{Z}_N$ factors that are spontaneously broken to
the trivial element of the group. The case of a large-$N$ SU($N$)
gauge theory with adjoint fermions on a compact domain has an even
richer phase structure~\cite{Hollowood:2009sy}.

If the three spatial directions have a finite extension and $N$ is
finite, no phase transition can occur at any size of the
system. However one can still investigate whether the distribution of
the Polyakov loop in a given spatial direction has a single peak in
zero (S, symmetric phase) or two separate peaks symmetric around zero
(A, asymmetric phase), keeping in mind that at infinite volume the
Polyakov loop distribution should display a single peak. The S-phase
is the finite volume and finite $N$ equivalent of the thermodynamic
phase in which the system is symmetric under the $\mathbb{Z}_N$
symmetry related to the direction wrapped by the Polyakov loop, while
the A-phase is the finite volume and finite $N$ equivalent of the
broken phase. We stress once again that at finite $N$ and on a finite
volume there are no distinct phases, but only different regimes. The
terminology here is used only for convenience.

We measure the distributions of the Polyakov loops in all the
directions. In all the lattices we consider, we always find a temporal
S-phase, which means that we are correctly simulating the confined
thermal phase of the infinite volume system. In lattices with geometry
$N_t \times N_s^3$, we are not interested in separating the equivalent
spatial directions. Hence we will say that the system is in a spatial
S-phase if all the spatial directions show a single peak distribution
centered in zero.

\begin{figure}[ht]
\centering
\includegraphics*[width=.8\textwidth]{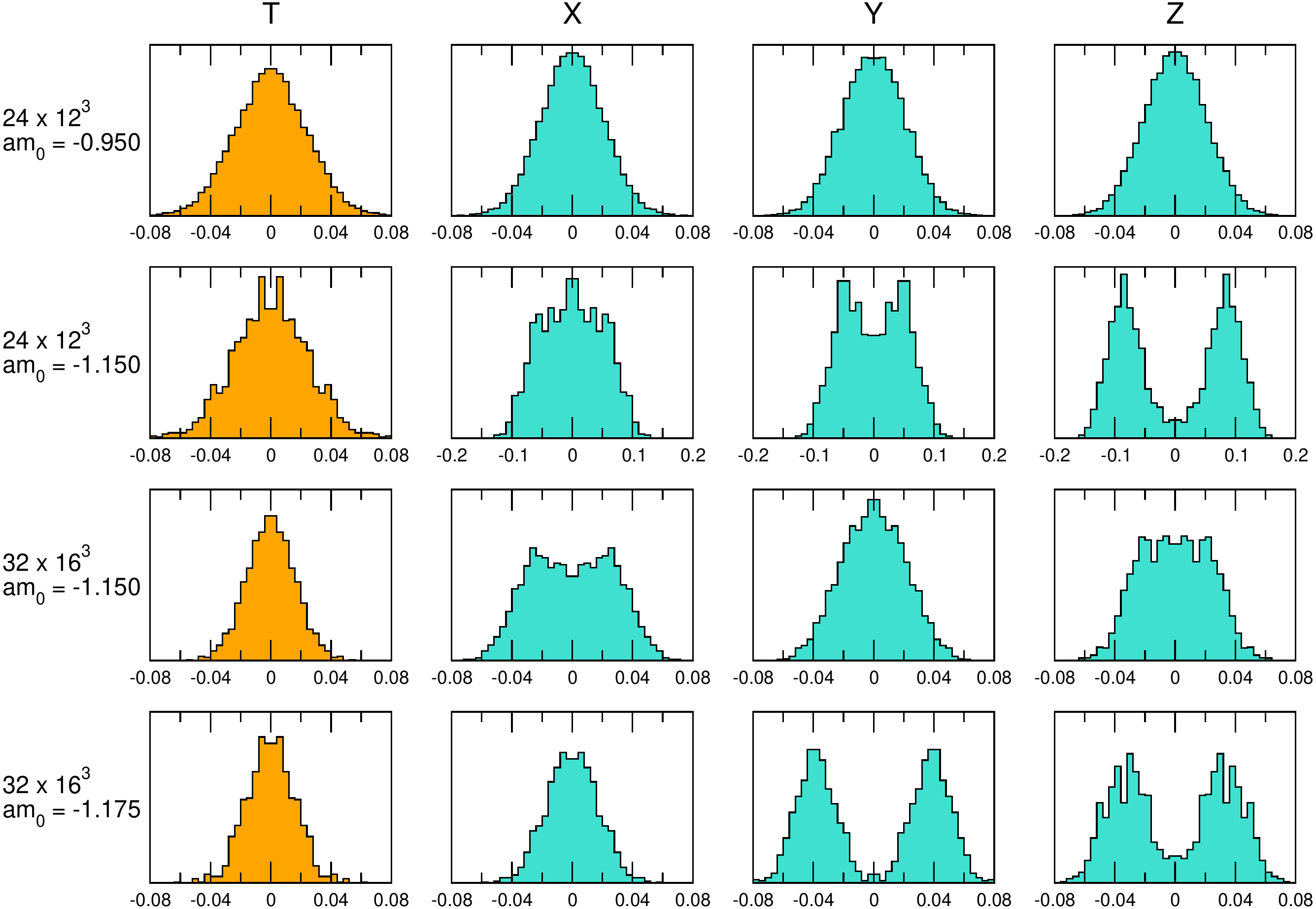}
\caption{Some distributions of Polyakov loops in the temporal and
  spatial directions. The distributions have been symmetrized by
  hand. At fixed volume, the distribution of the spatial Polyakov loop
  shows a single peak at zero at the higher mass, and it develops two
  peaks at the lower mass in some of the spatial directions. At fixed
  mass, it shows a single peak on the larger lattice, and it develops
  double peaks on the smaller lattice in some of the spatial
  directions.}
\label{fig:poly:1}
\end{figure}

\begin{figure}[ht]
\centering
\includegraphics*[width=.8\textwidth]{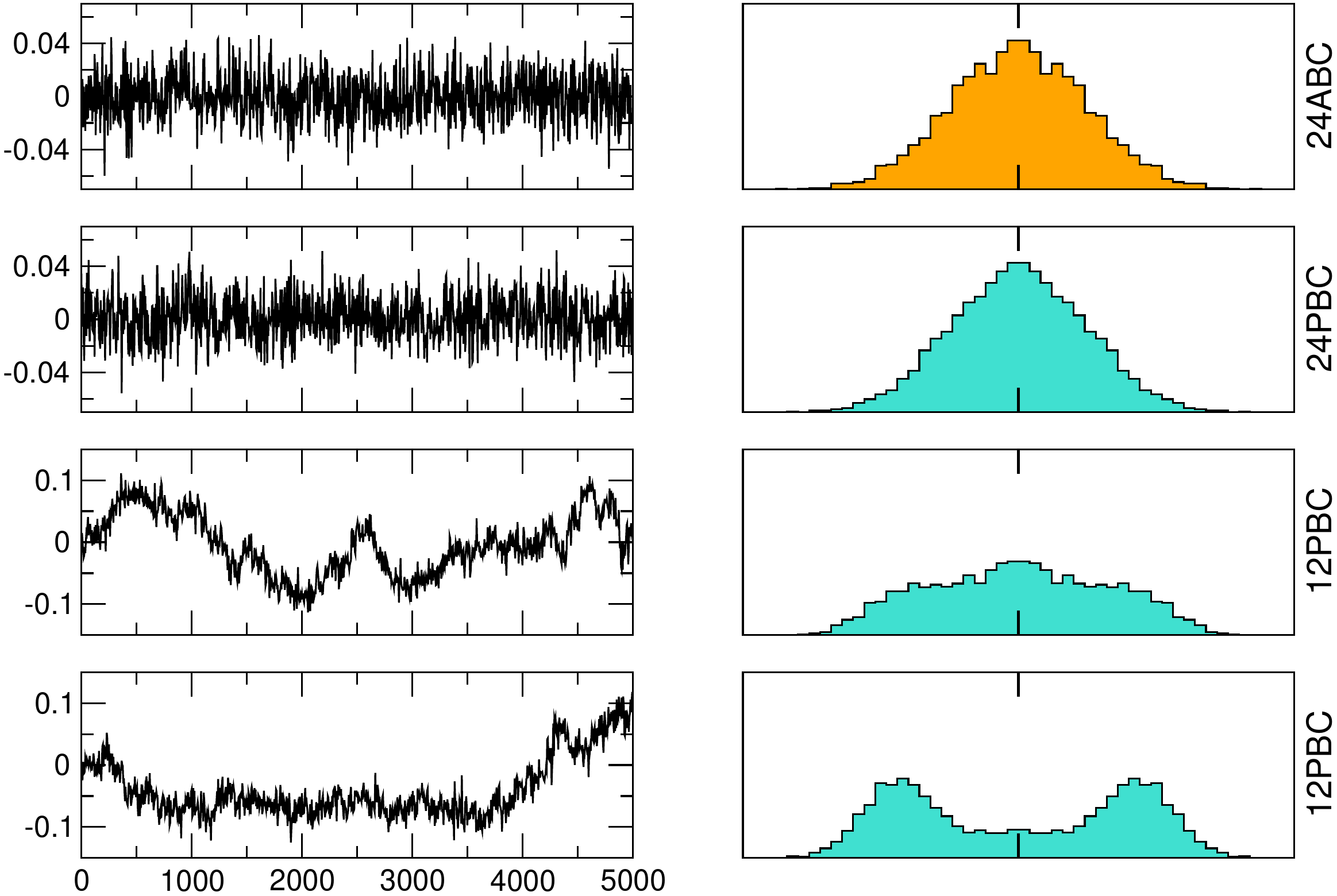}
\caption{Histories and distributions of Polyakov loops in the temporal
  (with fermionic antiperiodic boundary conditions, ABC) and spatial
  directions (with fermionic periodic boundary conditions, PBC), for
  bare mass $-1.125$ on the $24 \times 24 \times 12^2$ lattice. The
  distributions have been symmetrized by hand.}
\label{fig:poly:2}
\end{figure}

For the $16 \times 8^3$ lattice, we find the spatial A-phase below
$am_0\approx -0.975$. The crossover from the S-phase to the A-phase
is very smooth. A summary of the realization of the spatial center in
the $16 \times 8^3$ can be found in Table~\ref{tab:poly:8}. The two
maxima of the spatial distributions are better defined on the
$24\times 12^3$ lattice, for which the spatial A-phase is found below
$am_0=-1.05$ (see Table~\ref{tab:poly:12} for a summary). For the $32
\times 16^3$ lattice, all the simulated masses show distinct maxima in
the distribution of the spatial Polyakov loops, except at the higher
mass where we find a broad distribution with a flat region in the
middle (see Table~\ref{tab:poly:16} for a summary). We show some
distributions for the $24\times 12^3$ and $32 \times 16^3$ lattices in
Fig.~\ref{fig:poly:1}. We also simulated a $24\times24\times12^2$
lattice at $am_0=-1.125$. In this case the large spatial direction is
in the S-phase, while the smaller ones are in the A-phase. In
Fig.~\ref{fig:poly:2} we plot the histories and the distributions for
all the Polyakov loops for the $24\times24\times12^2$ lattice.

\begin{figure}[ht]
\centering
\includegraphics*[width=.6\textwidth]{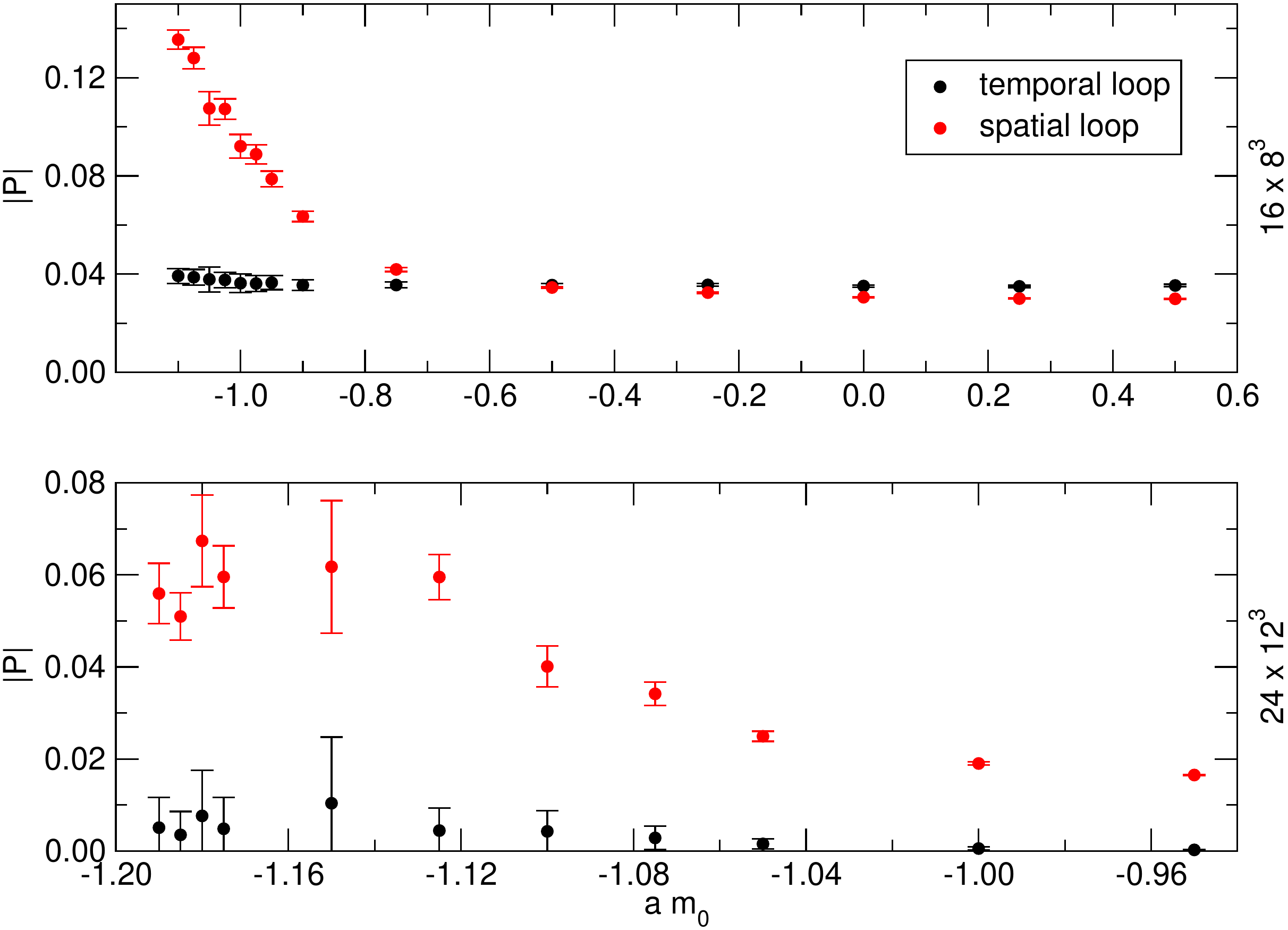}
\caption{Absolute value of the Polyakov loop on the $16 \times 8^3$
  and $24\times12^3$ lattices.}
\label{fig:abspoly}
\end{figure}

In Fig.~\ref{fig:abspoly}, we also report the absolute value of the
Polyakov loop on the $16 \times 8^3$ and $24\times12^3$ lattices. As
the mass is decreased, this quantity undergoes a crossover from a
lower value to a higher one. This crossover moves to a noticeably
lower mass when the lattice size is increasing, indicating that the
A-phase disappears in the large-volume limit.

The crossover from the S-phase to the A-phase as the spatial volume is
decreased is a clear indication that at least our S-phase is not in
the femtoworld regime, which is in the nearby of the zero-volume
limit, and which is a possible source of large systematic errors in a
lattice simulation~\cite{Fodor:2009wk}. Our data are actually
consistent with the picture that the S-phase is connected with the
infinite volume limit. However, as we shall show in detail, we still
have large finite-volume effects for several of the measured
observables.

\section{String tension from correlators of Polyakov loops}
\label{sect:lattice:string}
For a SU($N$) gauge theory in the confined phase, a static
quark-antiquark pair in the fundamental representation at large
separation $R$ is bound by the potential
\begin{equation}
V(R) = \sigma R \ ,
\end{equation}
where the string tension $\sigma$ is the dynamically generated scale
of the system. $\sigma$ is the string tension in an effective string
theory describing the low energy dynamics of confining flux tubes
connecting the quark and the antiquark. Contrary to the adjoint string
tension (and to the fundamental string tension in QCD), the
chromoelectric field between two fundamental sources in a gauge theory
with adjoint matter is not screened. Hence, the asymptotic fundamental
string tension is a well-defined quantity.

It is easy to prove analytically on the lattice that any gauge theory
has a non-zero string tension at strong coupling and large fermion
masses. The relevant question for the system under study is whether a
region in bare parameter space exists, which is analytically
connected with the continuum limit, and where the string tension is
zero in the massless limit, as it should be if the theory is
conformal.

The string tension can be extracted from correlators of Polyakov
loops. In particular, consider the plane defined by $x_0=0$ and a
fixed transverse coordinate in one arbitrary spatial direction
(e.g. we can consider the case of constant coordinate $x_1$ in the first
direction), and let us define:
\begin{equation}
  P_0(x_1) =
  \frac{1}{N_s^2} \sum_{x_2,x_3} \frac{1}{N}\Tr \left( \prod_{n_0=0}^{N_t - 1}
    U(an_0,{\mathbf r};0) \right) \ , 
\end{equation}
where ${\mathbf r}$ is the spatial vector with coordinates
$(x_1,x_2,x_3)$.
For the vacuum-subtracted correlator of this quantity one finds
\begin{equation}
  \label{eq:polycorr}
  \langle P_0(x_1)^{\dag} P_0(x_1+X_1) \rangle 
  - \left| \langle P_0(x_1) \rangle \right|^2 = 
  \sum_{n=0}^{\infty}  |c_n(L_t)|^2 e^{- E_n(L_t) X_1} \ , 
  \qquad E_n > 0 \ \forall n \ , 
\end{equation}
where the sum runs over all states $|n \rangle$ of the Hamiltonian
with non-vanishing overlap $c_n = \langle 0 | P_0(x_1) | n \rangle$. At
large $X_1$ the sum is dominated by the term $|c_0(L_t)|^2 e^{- E_0(L_t)
  X_1}$ associated with the exponential with the lowest decay rate,
$E_0(L_t)$, which is the energy of the groundstate in the fundamental
string sector. For a confining theory, up to subleading corrections
one finds~\cite{deForcrand:1984cz} 
\begin{equation}
\label{string_l}
E_0 (L_t) = \sigma L_t - c \pi(D-2)/(6 L_t) \ ,
\end{equation}
where $c$ depends on the number of massless fermionic and bosonic
modes propagating along the string and $D$ is the dimension of the
system ($D=4$ in our case). Analogously, if the Polyakov loop wraps a
spatial direction, its vacuum-subtracted zero momentum correlators in
the temporal direction define the so-called spatial string
tension. For a confined theory at zero temperature, the spatial and
the temporal string tensions coincide if the system is the S-phase.

In SU($N$) Yang-Mills theories, the effective theory describing the
large-distance dynamics of the confining flux tube is a bosonic string
theory~\cite{Lucini:2001nv,Necco:2001xg,Luscher:2002qv,Caselle:2004er}. 
General
arguments lead to the following expansion of $E_0$~\cite{Aharony:2009gg} 
\begin{equation}
  \label{string_ak}
  E_0 (L_t) = \sigma L_t - \frac{\pi(D-2)}{6 L_t} - \frac{1}{2} \left(
    \frac{\pi(D-2)}{6}\right)^2 \frac{1}{\sigma L_t^3} + \sigma L_t c_6 + \dots \ ,
\end{equation}
where $c_6$ is a term ${\cal O}((1/(\sigma^3 L_t^6))$. Note that
Eq.~(\ref{string_ak}) is the truncation to second order in $1/(\sigma
L_t^2)$ of the ground state energy of a bosonic string of the
Nambu-Goto type:
\begin{equation}
\label{string_ng}
E_0 (L_t) = \sigma L_t\, \sqrt{ 1 - \frac{\pi(D-2)}{3 \sigma L_t^2}} \ .
\end{equation}
This might suggests that the effective string theory describing the
large-distance dynamics of the confining flux tube is actually
Nambu-Goto. These considerations do not generalize immediately to the
case of dynamical fermions in the adjoint representation. In fact
differently from SU($N$) pure Yang-Mills, in this case it is possible
to construct explicitly fermionic open-string states. Hence, it could
be that the effective theory is not bosonic, in which case the firmest
result available is given by Eq.~(\ref{string_l}), with $c$ (unknown
{\em a priori}) counting the zero-modes of the effective string. In
our analysis, we will assume that the effective theory is bosonic and
the string tension will be obtained from correlator of Polyakov loops
assuming Eq.~\eqref{string_ng}. This assumption is justified {\em a
  posteriori} by two crucial observations: first, the string tension
obtained in this way is in complete agreement with the static
potentials and forces computed by Wilson loops
(Sect.~\ref{sect:lattice:wilson}), for which no effective string
theory is assumed; in addition, the low-energy dynamics will be found
to be an effective Yang-Mills with small corrections (accordingly to
the discussion in Sect.~\ref{subsect:locking}) for all the simulated
masses, which shows that the choice of Eq.~\eqref{string_ng} is
self-consistent.

For a theory that confines at zero temperature and undergoes a
deconfinement phase transition at some critical temperature, the
correlator in Eq.~(\ref{eq:polycorr}) still decays exponentially with
the distance in the deconfined phase, provided the sources are
screened; however, the inverse of the corresponding energy is now
associated with a screening length, the Debye screening length. Hence,
the exponential decay of Polyakov loop correlators by itself does not
imply the existence of a string tension. In order to see that the
theory is confining, the validity of Eq.~(\ref{string_ng}) as $L_t$ is
varied needs to be proved. Alternatively, one has to show that the
static potential~(Sect.~\ref{sect:lattice:wilson}) is asymptotically
linear and the slope of the linear part is related to the ground state
mass extracted from Polyakov loop correlators via
Eq.~(\ref{string_ng}).

In general, extracting numerically the string tension from the
correlator~(\ref{eq:polycorr}) proves to be technically hard, since
the signal-over-noise ratio decays exponentially with the
separation. A good degree of success is achieved if the zero-momentum
Polyakov line is replaced by a fuzzy operator, and a reliable signal
can be obtained if a variational procedure that involves different
fuzzy operators is set up. There are several ways of achieving this;
here we follow Refs.~\cite{Lucini:2004my,Lucini:2004eq}. In practice,
a recursive procedure is implemented, which allows to obtain smeared
links at step $l+1$ from the fuzzy links at $l$ via the relationship
\begin{equation}
\label{smearing}
U^{(l+1)} (x,\mu) = \mbox{Proj}\left (U^{(l)}(x,\mu) + \alpha
  S^{(l)}(x,\mu) 
  + \delta D^{(l)}(x,\mu) \right) \ ,
\end{equation}
where $S^{(l)}(x,\mu)$ is the sum of the four length three
non-backtracking lattice paths from $x$ to $x + a\hat{\mu}$ ({\em
  staple}) and $D^{(l)}(x,\mu)$ is the sum of the 16 length-five
non-backtracking paths with the same start and end points (only the
directions that are orthogonal to the direction in which correlations
are taken enter the sums). The constants $\alpha$ and $\delta$ are are
fixed empirically in such a way that the signal is optimal. Since the
(weighted) sum of the paths is not an element of the group, to obtain
an object that can be interpreted as a fuzzy link this sum needs to be
reprojected onto SU($N$); this is the meaning of the operator Proj$()$
in Eq.~(\ref{smearing}). After $k$ steps of smearing, consecutive
pairs of smeared links going in the same direction can be multiplied
to produce blocked links. The combination of smearing and blocking
yields the link set $\{U^{(b)}(x,\mu)\}$ at blocking level $b$,
which can be used to compute the fuzzy Polyakov loop operator
\begin{equation}
P_0^{(b)}(x) = \frac{1}{N_s^2} \sum_{y,z} \frac{1}{N}\Tr 
\left( \prod_{n_0=0}^{N_t/b} U^{(b)}(an_0,{\mathbf r};0) \right) \ ,
\end{equation}
and analogously for the other directions. For sake of definiteness, we
discuss the case of Polyakov loops winding in time, but similar
conclusions hold for Polyakov loops wrapping around the other
directions. The element $bc$ of the correlation matrix $\tilde{C}(X_1)$
is then defined as
\begin{equation}
\label{eq:corrmatr}
\tilde{C}_{bc}(X_1) = \langle P_0^{(b)}(x_1)^{\dag} P_0^{(c)}(x_1+X_1) \rangle
- \langle P_0^{(b)}(x_1) \rangle^{\dag} \langle P_0^{(c)}(x_1)  \rangle \ .
\end{equation}
As functions of $X_1$, diagonal correlators involving the eigenvectors
associated with the largest eigenvalues of $\tilde{C}^{-1}(0)
\tilde{C}(1)$ decay as single exponentials with energies $E_0,E_1,\dots$
already at distances of a few lattice spacings. The negligible
contamination from excited states eliminates the need to go to large
distances to identify the stringy state with the lowest energy; at the
same time, this procedure provides an estimate of energies of excited
states, associated to the single-exponential behavior of diagonal
correlators of eigenvectors corresponding to smaller eigenvalues,
although the reliability of the excited spectrum depends on how large
the variational basis is. The efficiency of the variational procedure
is manifest in the overlap of the vacuum with the lowest-lying stringy
state (i.e. the equivalent of the coefficient $|c_0|^2$ in
Eq.~(\ref{eq:polycorr})), which would be one in the ideal case in
which the variational procedure identified the exact state we are
interested in. In the considered fuzzying scheme, with a careful
choice of the parameters $\alpha$ and $\delta$ it is possible to reach
overlaps of the order of 0.9-0.95, which makes the
contribution of excited states negligible already at distances of the
order of two lattice spacings. Physically,
the process of blocking and smearing allows us to build variational
trial states on the scale of physical distances, while simple Polyakov
loop correlators probe the physics on the scale of the lattice
spacing, which is sensitive to ultraviolet fluctuations.

\begin{figure}[ht]
\centering
\includegraphics*[width=.6\textwidth]{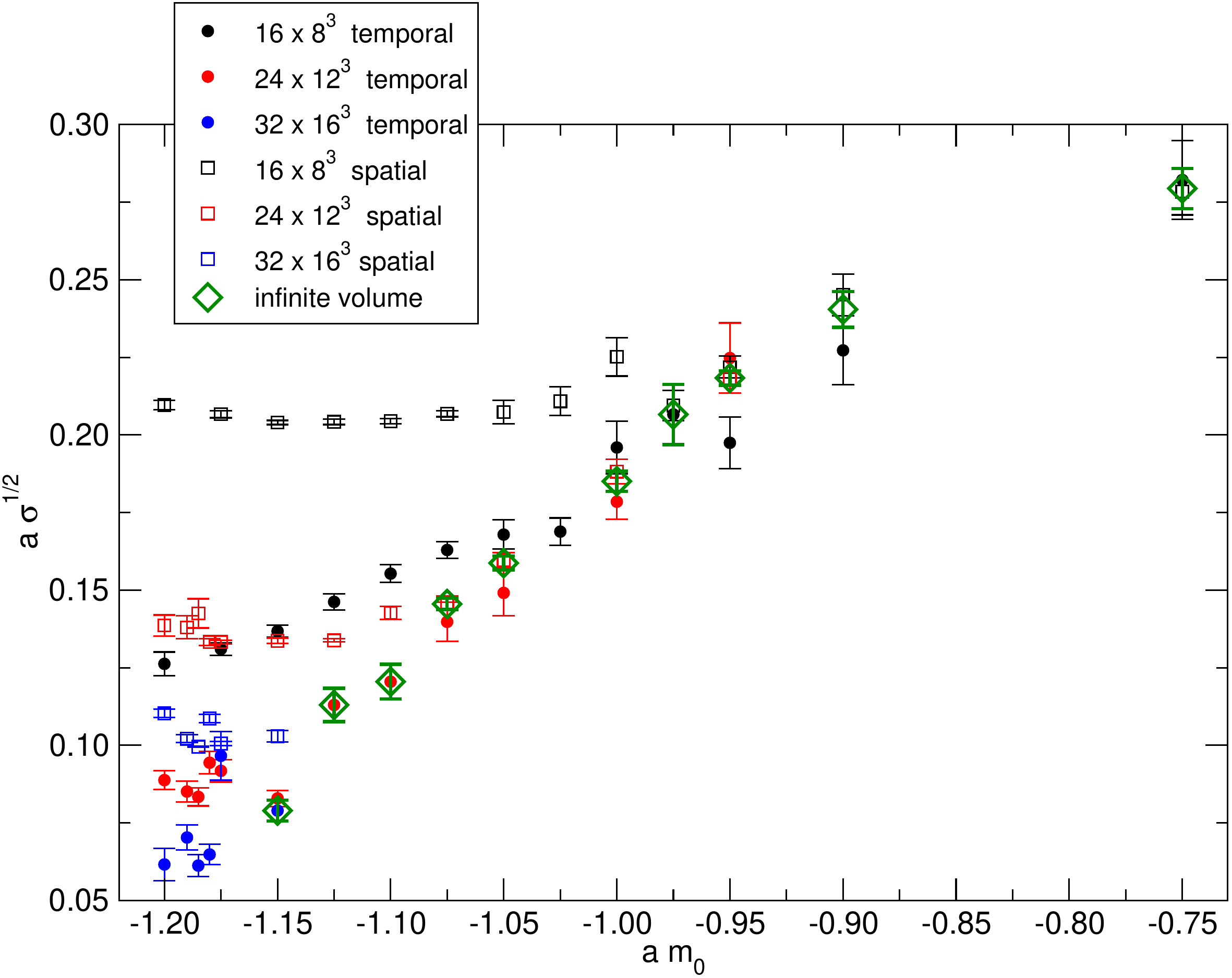}
\caption{Spatial and temporal string tensions at various lattice sizes
  as a function of $am_0$. Also shown is the infinite volume estimate.}
\label{fig:sigma}
\end{figure}

We remind that we use Eq.~(\ref{string_ng}) for extracting the string
tension from $E_0$ \footnote{Alternatively, Eq.~(\ref{string_ak}) can
  be used, since it gives results that are compatible well within
  errors with Eq.~(\ref{string_ng}); on the contrary,
  Eq.~(\ref{string_l}) gives discrepancies of up to 20\% for loops
  winding the spatial directions.}.  Our results for the string
tension are plotted in Fig.~\ref{fig:sigma}. For the smearing and
blocking procedures, we have used $\alpha = 0.4$ and $\delta =
0.16$. As expected, at high mass the spatial and the temporal string
tensions agree and are independent of the lattice size. The signal
provided by the Polyakov loop correlator is clean and the overlap
$|c_0|^2$ is of order $0.9-0.95$. As $am_0$ is reduced, on the smaller
lattices the spatial and the temporal string tensions depart, and the
overlap becomes of order $0.8$. The disagreement between the temporal
and the spatial string tensions is an indication that finite size
effects are starting to play a major role. As a matter of fact, if we
simulate at fixed values of $am_0$ on increasingly larger lattices, we
find that the spatial and the temporal string tensions eventually
agree, and that they also agree with the temporal string tension of
the smaller lattices. This shows that the temporal string tension is
less affected by finite size effects. This is hardly surprising, since
the onset of the departure between the spatial and temporal string
tension arises where the system goes from the S- to the A-phase. This
also confirms that in the A-regime strictly speaking it is not correct
to talk about a spatial string tension, since the mass of the Polyakov
loop is not associated to confining strings. However, for convenience
we shall still use the string language. For a given volume, if the
mass is lowered below the onset of spatial deconfinement, the
agreement of the temporal string tension and the string tensions§
computed on larger volumes is lost, and the former flattens out. This
kind of finite size effects appears when the correlation length
associated to the string tension becomes of the order of the spatial
lattice size. In fact, for the plateau values we find $a^2 \sigma_t
\simeq 4 N_t^{-2} = N_s^{-2}$. Taken at face value, this would imply
that in the thermodynamic and chiral limit $a \sqrt{\sigma_t} = 0$. However,
since these results have been obtained in a phase were finite size
artefacts play a major role, a confirmation of this statement on
larger lattices is necessary before we can conclude that there is no
asymptotic string tension in the massless limit.

We estimate the string tension at infinite volume by choosing the
determinations that are reasonably safe from finite volume
effects. Our strategy is based on the following observations. At fixed
mass and in the infinite volume limit, the temporal and spatial string
tensions must coincide. At fixed volume we observe that at large
enough mass, the temporal and spatial string tensions coincide. In
general the spatial string tension can be determined more accurately,
since correlators of shorter Polyakov loops have smaller relative
errors. On the other hand, the temporal string tension is less
affected by finite volume effects and in particular is always well
defined. Whenever the spatial and temporal string tensions agree
within one standard deviation, we consider the weighted average of the
two:
\begin{gather}
  \sqrt{\sigma} = \frac{ \frac{\sqrt{\sigma_s}}{(\Delta \sqrt{\sigma_s})^2} + \frac{\sqrt{\sigma_t}} {(\Delta \sqrt{\sigma_t})^2} }{ \frac{1}{(\Delta \sqrt{\sigma_s})^2} + \frac{1} {(\Delta \sqrt{\sigma_t})^2} } \ , \\
  \Delta \sqrt{\sigma} = \frac{1}{ \sqrt{ \frac{1}{(\Delta
        \sqrt{\sigma_s})^2} + \frac{1} {(\Delta \sqrt{\sigma_t})^2} }
  } \ .
\end{gather}
In the A-phase, where the temporal and spatial string tensions do not
agree anymore, the correct string tension to be considered is the
temporal one. At those values of the mass, for which more than one
volume is available, the result on the largest volume has been
considered.

More in detail, at the mass $am_0=-1$ both the $24\times 12^3$ and $16
\times 8^3$ lattices are available, and the two temporal string
tensions (plus the spatial string tension on the larger volume) are in
agreement at the $1\sigma$ level. For this mass and larger ones we
therefore expect that the temporal string tension, as determined on
the $24\times 12^3$ and $16 \times 8^3$ lattices, is affected by
smaller finite-volume effects than the statistical indetermination. In
particular, the string tensions on the $24\times 12^3$ lattice at
masses $am_0=-1$ and $-0.95$, and on the $16 \times 8^3$ lattice at
masses $am_0=-0.975, -0.9, -0.75, -0.5, -0.25, 0, 0.25$ are good
estimates for the infinite volume limit.

Moreover, at the mass $am_0=-1.15$ both the $32\times 16^3$ and
$24\times 12^3$ lattices are available, and the two temporal string
tensions agree at the $1\sigma$ level. Repeating the argument above,
the string tensions on the $32\times 16^3$ lattice at mass
$am_0=-1.15$, and on the $24\times 12^3$ lattice at masses $am_0=-1.125,
-1.1, -1.075, -1.05$ are good estimates of the infinite volume
limit. Below mass $am_0=-1.15$ we have no control on the finite volume
effects, so we just discarded those points.

The choices described above, and the results for the estimate of the
string tension at infinite volume are summarized in
Table~\ref{tab:inf_sigma}. All the measured string tensions and the
infinite volume estimates are also plotted in Fig.~\ref{fig:sigma}.

\begin{table}[ht]
\centering
\begin{tabular}{c|c|c|c}
\hline
$-am_0$ & inf. volume estimate of $a\sigma^{1/2}$ & method & static potential \& force \\
\hline
\hline
-0.25 & 0.4053(51) & w.a. on $16\times 8^3$ (S1) data       & --                       \\
0    & 0.3976(51) & w.a. on $16\times 8^3$ (S2) data       & --                       \\
0.25  & 0.352(11)  & w.a. on $16\times 8^3$ (S3) data       & --                       \\
0.5   & 0.3561(42) & w.a. on $16\times 8^3$ (S4) data       & --                       \\
0.75  & 0.2794(63) & w.a. on $16\times 8^3$ (S5) data       & --                       \\
0.9   & 0.2405(58) & w.a. on $16\times 8^3$ (S6) data       & --                       \\
0.95  & 0.2184(23) & w.a. on $24\times 12^3$ (B0) data      & $24\times 12^3$          \\
0.975 & 0.2066(97) & temporal s.t. on $16\times 8^3$ (A1)   & --                       \\
1     & 0.1851(33) & w.a. on $24\times 12^3$ (B1) data      & $24\times 12^3$          \\
1.05  & 0.1587(23) & w.a. on $24\times 12^3$ (B2) data      & $24\times 12^3$          \\
1.075 & 0.1455(19) & w.a. on $24\times 12^3$ (B3) data      & $24\times 12^3$          \\
1.1   & 0.1205(56) & temporal s.t. on $24\times 12^3$ (B4)  & $24\times 12^3$          \\
1.125 & 0.1130(54) & temporal s.t. on $24\times 12^3$ (B5)  & $24\times 24\times 12^3$ \\
1.15  & 0.0790(34) & temporal s.t. on $32\times 16^3$ (C0)  & $32\times 16^3$          \\
\hline
\end{tabular}
\caption{This table shows the infinite volume estimate for the string 
  tension (2nd column). Where different volumes are available for the 
  same bare mass, the larger one has been chosen. If the temporal and 
  spatial string tensions agree within one standard deviation, a weighted
  average (w.a.) between the two has been chosen, as explained in the text. 
  Otherwise the temporal string tension (s.t.) has been chosen. For
  the 
  bare mass $am_0 = -1$, the temporal string tensions on the $16\times
  8^3$ 
  and $24\times 12^3$ lattices agree at the $1\sigma$ level: this 
  suggests that for all the string tensions computed at this and
  higher 
  masses on the $16\times 8^3$ lattice, finite volume effects are 
  negligible. Also the temporal string tension computed on the 
  $32\times 16^3$ lattice at $am_0 = -1.15$ coincides at $1\sigma$ with 
  the one computed on the $24\times 12^3$ lattice.}
\label{tab:inf_sigma}
\end{table}

\section{Static force and potential from Wilson loops}
\label{sect:lattice:wilson}
A different way to compute the string tension is via the expectation
values of Wilson loops. The advantage of this method is that it can
show whether the mass extracted via Polyakov loop correlators is in
fact related to the existence of an asymptotic string tension at fixed
lattice geometry. As we will see, the disadvantage is that the
numerical results for the string tension extracted from Wilson loops
generally have larger statistical errors.

We consider the generic off-axis Wilson loop $W(T,\mathbf{R})$,
describing a quark-antiquark pair separated by a spatial distance
$\mathbf{R}=(X,Y,Z)$ and propagating in straight line in the temporal
direction. It will be useful to schematically decompose the close
parallel transport in its spatial $U_s[t;\mathbf{r} \to
\mathbf{r}+\mathbf{R}]$ and temporal $U_t[t \to t+T;\mathbf{r}]$
components:
\begin{equation}
W(T,\mathbf{R}) = \Tr \{
U_t[t \to t+T; \mathbf{r}]^\dagger
U_s[t; \mathbf{r} \to \mathbf{r}+\mathbf{R}]
U_t[t \to t+T; \mathbf{r}+\mathbf{R}]
U_s[t+T; \mathbf{r} \to \mathbf{r}+\mathbf{R}]^\dagger
\} \ .
\end{equation}
The off-axis component $U_s[t; \mathbf{r} \to \mathbf{r}+\mathbf{R}]$
is computed following the algorithm in Ref.~\cite{Bolder:2000un}. In
order to reduce the noise in the static potential, we build the Wilson
loops with smeared link variables. We choose a single step of HYP
smearing~\cite{Hasenfratz:2001hp}; the smeared link variable is a
function of all the links belonging to the unit hypercubes adjacent to
the original link. We found the HYP smearing effective enough for our
purposes, even not implementing a variational method. The main
disadvantage is that it deforms the static potential at short
distances; however it does not affect the determination of the string
tension, which is our main goal.

In the Hamiltonian gauge ($U(x,0)=1$), the expectation value of
$W(T,\mathbf{R})$ becomes the time-correlator for the operator
$M(\mathbf{R}) = \bar{Q}(\mathbf{r}) U_s[\mathbf{r} \to
\mathbf{r}+\mathbf{R}] Q(\mathbf{r}+\mathbf{R})$ which creates a heavy
quark-antiquark pair connected by a string:
\begin{eqnarray}
  \langle W(T,\mathbf{R}) \rangle &=&
  \frac{1}{Z(L_t)} \Tr [ e^{-(L_t-T)H} M(\mathbf{R})^\dagger 
  e^{-T H} M(\mathbf{R}) ] = \nonumber \\
  &=& \frac{1}{Z(L_t)} \sum_{nm} \left| \langle n,\mathbf{R} | 
    M(\mathbf{R}) | m \rangle \right|^2 
  e^{-L_t E_m} e^{-T \left[ V_n(R) - E_m \right] } \ ,
\end{eqnarray}
where we have inserted the gauge invariant states $|m\rangle$ with
energy $E_m$, and the states in presence of a quark-antiquark pair
$|n,\mathbf{R}\rangle$ with energy $V_n(R)$. The energy $V_0(R)$ of
the fundamental state in presence of the quark-antiquark pair is the
static potential. $Z(L_t)$ is the partition function, which can be
decomposed in terms of the gauge invariant states $|m\rangle$:
\begin{equation}
Z(L_t) = \Tr e^{-L_t H} = \sum_m e^{-L_t E_m} \ .
\end{equation}
If the temporal extension $L_t$ is large enough, we can identify three
regimes for the Wilson loops while $T$ changes:
\begin{enumerate}
\item Small values of $T$ with respect to the temporal extension ($T
  \ll L_t$). This is the usual zero temperature limit. In this case
  only the vacuum survives among the $|m\rangle$ states:
  \begin{equation}
    \langle W(T,\mathbf{R}) \rangle \simeq \sum_{n} |\alpha_n(R)|^2 e^{-T [V_n(R) - E_0 ] } \ , \qquad \textrm{with } \sum_{n} |\alpha_n(R)|^2 = 2  \ ,
  \end{equation}
  where the relationship for the coefficients $\alpha_n$ comes from
  $\langle W(0,\mathbf{R}) \rangle = 2$. It is interesting to notice
  that, since $\langle W(T,\mathbf{R}) \rangle \le 2$ for each value
  of $T$, $V_n(R)$ must be larger that the vacuum energy $E_0$ for
  each value of $R>0$ (because of the lattice discretization, the
  potential is bounded from below). Therefore in this regime, the
  Wilson loop is decreasing in $T$ at every fixed $\mathbf{R}$.
\item Values of $T$ comparable with the temporal extension ($L_t-T \ll
  L_t$). In this case only the $|0,\mathbf{R}\rangle$ survives among
  the states with external charges:
  \begin{equation}
    \langle W(T,\mathbf{R}) \rangle \simeq e^{-L_t [V_0(R) - E_0 ]} 
    \sum_{m} |\beta_m(R)|^2 e^{-(L_t-T) \left[ E_m - V_0(R) \right] } \ .
  \end{equation}
  Assume that the sum is dominated by some $m$. If the quantity $E_m -
  V_0(R)$ is positive, the Wilson loop at fixed $\mathbf{R}$ is
  increasing as $T$ approaches $L_t$, towards its extremal value
  $\langle W(L_t,\mathbf{R}) \rangle \simeq \sum_{m} |\beta_m(R)|^2
  \le 2 $.  If the quantity $E_m - V_0(R)$ is negative, the Wilson
  loop at fixed $\mathbf{R}$ is decreasing towards its extremal
  value. Notice that, since the states propagating around the torus
  are not the same as the ones propagating inside the Wilson loop, the
  Wilson loop is not symmetric in $T$ around $L_t/2$ as it usually
  happens for other correlators.
\item Intermediate values of $T$ ($a \ll T \ll L_t$). In this case the
  Wilson loop reduces to a single exponential:
  \begin{equation}
    \langle W(T,\mathbf{R}) \rangle \simeq A(\mathbf{R}) e^{-T [V_0(R) - E_0 ] } \ .
  \end{equation}
  This is the useful regime which we will try to identify in our
  numerical simulations to extract the static potential.
\end{enumerate}

The first and second regions are always visible if the temporal length
$T$ of the Wilson loop is too small or too large. With the extra
difficulty that the Wilson loops are not symmetric in $T \to L_t-T$,
the computational problem for the static potential is similar to the
one for other correlators: it is important to have a large enough
lattice in such a way that the third region opens up in the middle. An
effective potential is therefore defined, with the property that it
shows a plateau in the third region (if visible), and the value of the
plateau is actually the static potential. We use two different methods
to extract the effective potential from the Wilson loops.
\begin{description}
\item[Potential1]. The easiest method consists in defining an effective potential as:
\begin{equation}
\label{eq:eff_potential}
V_\mathrm{eff}(T,\mathbf{R}) = - \frac{1}{a} \log \frac{\langle W(T+a,\mathbf{R}) \rangle}{\langle W(T,\mathbf{R}) \rangle} \ .
\end{equation}
If we can see a plateau in the effective potential as a function of
$T$, it means that we can isolate the single-exponential region. The
value $V(\mathbf{R})$ is then extracted by fitting the plateau of the
effective potential with a constant. Notice that since the lattice
breaks rotational invariance, we consider the potential as a function
of $\mathbf{R}$, and not of its module only. An unbiased estimate for
the average of the potential, and an estimate of its error are
obtained by applying Eq.~\eqref{eq:eff_potential} to a set of
bootstrap ensembles.
\item[Potential2]. When the single-exponential region is not visible,
  we use the Prony's method~\cite{Fleming:2009wb} for taking into
  account also the first excited state. We refer to the literature for
  the general idea, while we summarize here the used formulae. Having
  chosen a value of $\mathbf{R}$, for every value of $T$ we solve the
  following second-order equation:
  \begin{flalign}
    &\left[ W(T,\mathbf{R})W(T+2a,\mathbf{R}) - W(T+a,\mathbf{R})^2 \right] x^2 + \\
    & + \left[ W(T+a,\mathbf{R})W(T+2a,\mathbf{R}) - W(T,\mathbf{R})W(T+3a,\mathbf{R}) \right] x + \\
    & + \left[ W(T+a,\mathbf{R})W(T+3a,\mathbf{R}) - W(T+2a,\mathbf{R})^2 \right] = 0 \ .
  \end{flalign}
  If $x_0$ is the largest solution (but smaller than $1$), we define
  an effective potential as
  \begin{equation}
    V_\mathrm{eff}(T,\mathbf{R}) = - \frac{1}{a} \log x_0 \ .
  \end{equation}
  As for the previous method, the value $V(\mathbf{R})$ is extracted
  by fitting the plateau of the effective potential with a
  constant. The whole procedure is implemented via a bootstrap, in
  order to get an unbiased estimate for the average of the energies,
  and an estimate of its error.
\end{description}

Whenever we can compute the potential with both methods we observe
that they always give compatible results, but slightly smaller errors
and better determinations of the plateaux are obtained with the method
\textbf{Potential2}. Even though the string tension can be in
principle extracted from a linear fit of the static potential in a
large distance region, it is instructive to determine it also from the
force $F = - \frac{dV}{dR}$. At large $R$, $F \sim \sigma$. For the
determination of $F$, we use the following methods.
\begin{description}
\item[Force1]. We use generalized Creutz ratios to define an effective
  force with off-axis Wilson loops:
  \begin{equation}
    F_\mathrm{eff}(T,\mathbf{R},a\mathbf{n}) = 
    - \frac{1}{a^2 |\mathbf{n}|} \log \frac{
      \langle W(T+a,\mathbf{R}+a\mathbf{n}) \rangle 
      \langle W(T,\mathbf{R}) \rangle }{
      \langle W(T,\mathbf{R}+a\mathbf{n}) \rangle 
      \langle W(T+a,\mathbf{R}) \rangle } \ .
  \end{equation}
  We use $\mathbf{n}$ vectors of the form $(1,0,0)$, $(1,1,0)$ and
  $(1,1,1)$ and permutations. We identify the plateau of the effective
  force as a function of $T$, and we fit it with a constant
  $F(R_I)$. The statistical error is determined by a bootstrap
  procedure. The improved distance $R_I$ is defined as in
  Ref.~\cite{Sommer:1993ce} to be:
  \begin{equation}
    R_I= \left| 4\pi \frac{G(\mathbf{R}+\mathbf{n})-G(\mathbf{R})}{a |\mathbf{n}|} \right|^{-1/2} \ ,
  \end{equation}
  where $G(\mathbf{R})$ is the three-dimensional free-scalar
  propagator on the lattice.
\item[Force2]. Plateaux in the Creutz ratios are visible only in a
  region where only a single exponent dominates in the expansion of
  the Wilson loop. In most of the cases we need to take into account
  the first excited state. An effective force can be defined by
  using the effective potentials computed by the method
  \textbf{Potential2}:
  \begin{equation}
    F_\mathrm{eff}(T,\mathbf{R},a\mathbf{n}) = 
    - \frac{V_\mathrm{eff}(T,\mathbf{R}+a\mathbf{n}) 
      - V_\mathrm{eff}(T,\mathbf{R})}{a |\mathbf{n}|} \ .
  \end{equation}
  We identify the plateau of the effective force as a function of
  $T$, and we fit it with a constant $F(R_I)$. Expectation value and
  error of $F(R_I)$ are estimated by means of a bootstrap procedure.
\end{description}

\begin{figure}[ht]
\centering
\includegraphics*[width=.6\textwidth]{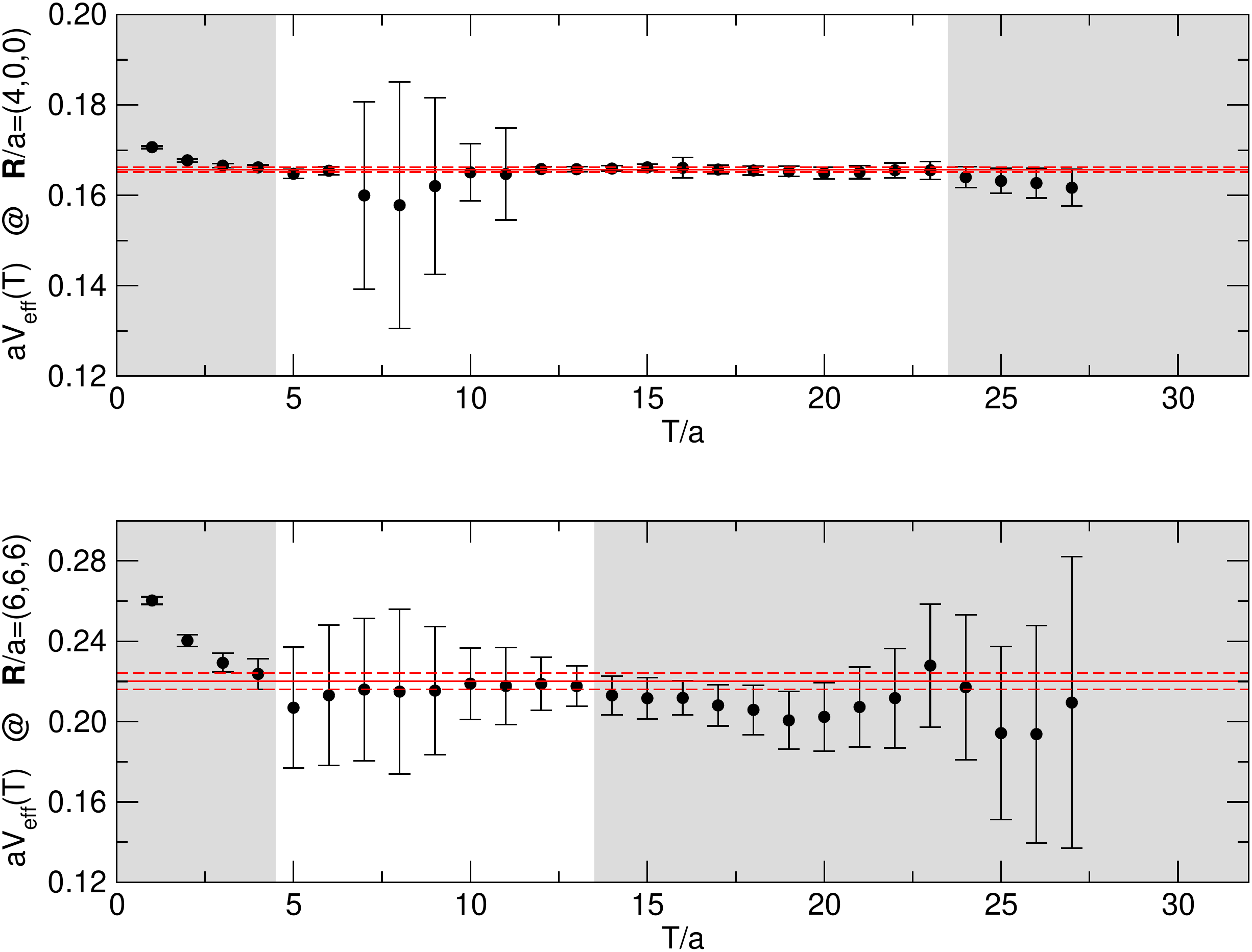}
\caption{Effective potentials at $\mathbf{R}=(4a,0,0)$ and
  $\mathbf{R}=(6a,6a,6a)$ computed on the $32\times 16^3$ lattice at bare
  mass $am_0=-1.15$. Plateaux have been chosen in the white
  regions. The red lines represent the values of the potential (with
  errors), obtained by fitting the effective potential with a constant
  in the white region. Errors have been computed with a bootstrap
  procedure.}
\label{fig:plateaux}
\end{figure}

The static potentials presented in this section have been computed
with the method \textbf{Potential2}. In Fig.~\ref{fig:plateaux}, two
typical effective potentials are shown, together with the fit range
and the result of the constant fit.

\begin{figure}[ht]
\centering
\includegraphics*[width=.6\textwidth]{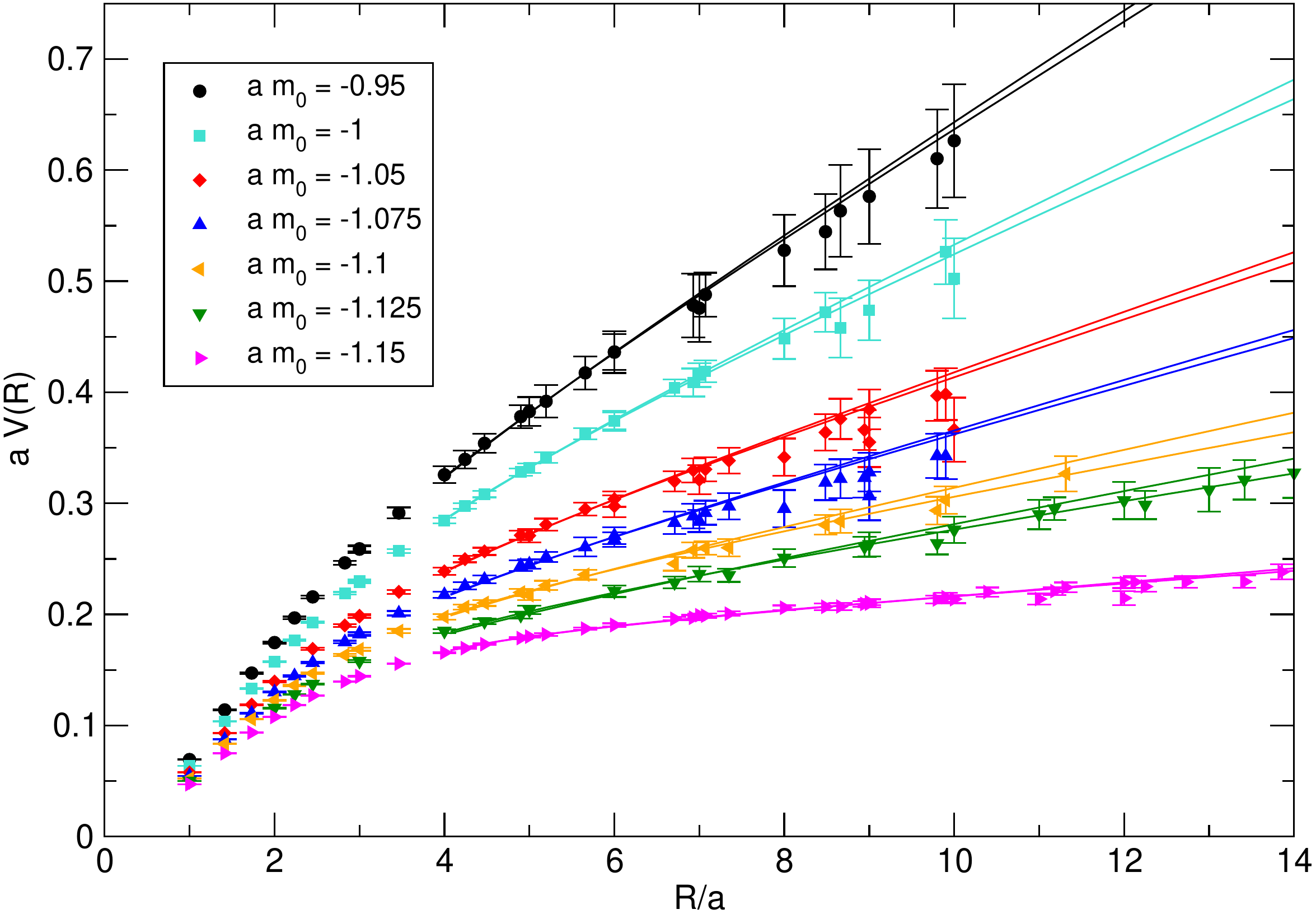}
\caption{Static potentials computed with the method
  \textbf{Potential2} (see Sec.~\ref{sect:lattice:wilson}) from Wilson
  loops with HYP smearing. Data for bare masses $-0.95$, $-1$,
  $-1.05$, $-1.075$, $-1.1$ are obtained on a $24 \times 12^3$
  lattice; data for bare mass $-1.125$ are obtained on a $24 \times 24
  \times 12^2$ lattice; data for bare mass $-1.15$ are obtained on a
  $32 \times 16^3$ lattice. Superimposed is the function $V(R)=\sigma
  R + \mu + c/R$, where $\sigma$ has been taken from
  Table~\ref{tab:inf_sigma} (the two curves correspond to $\sigma -
  \Delta \sigma$ and $\sigma + \Delta \sigma$), while $\mu$ and $c$
  have been obtained with a fit in the region $R \ge 3a$.}
\label{fig:allpotentials}
\end{figure}

\begin{figure}[ht]
\centering
\parbox{.5\textwidth}{
\includegraphics*[width=.4\textwidth]{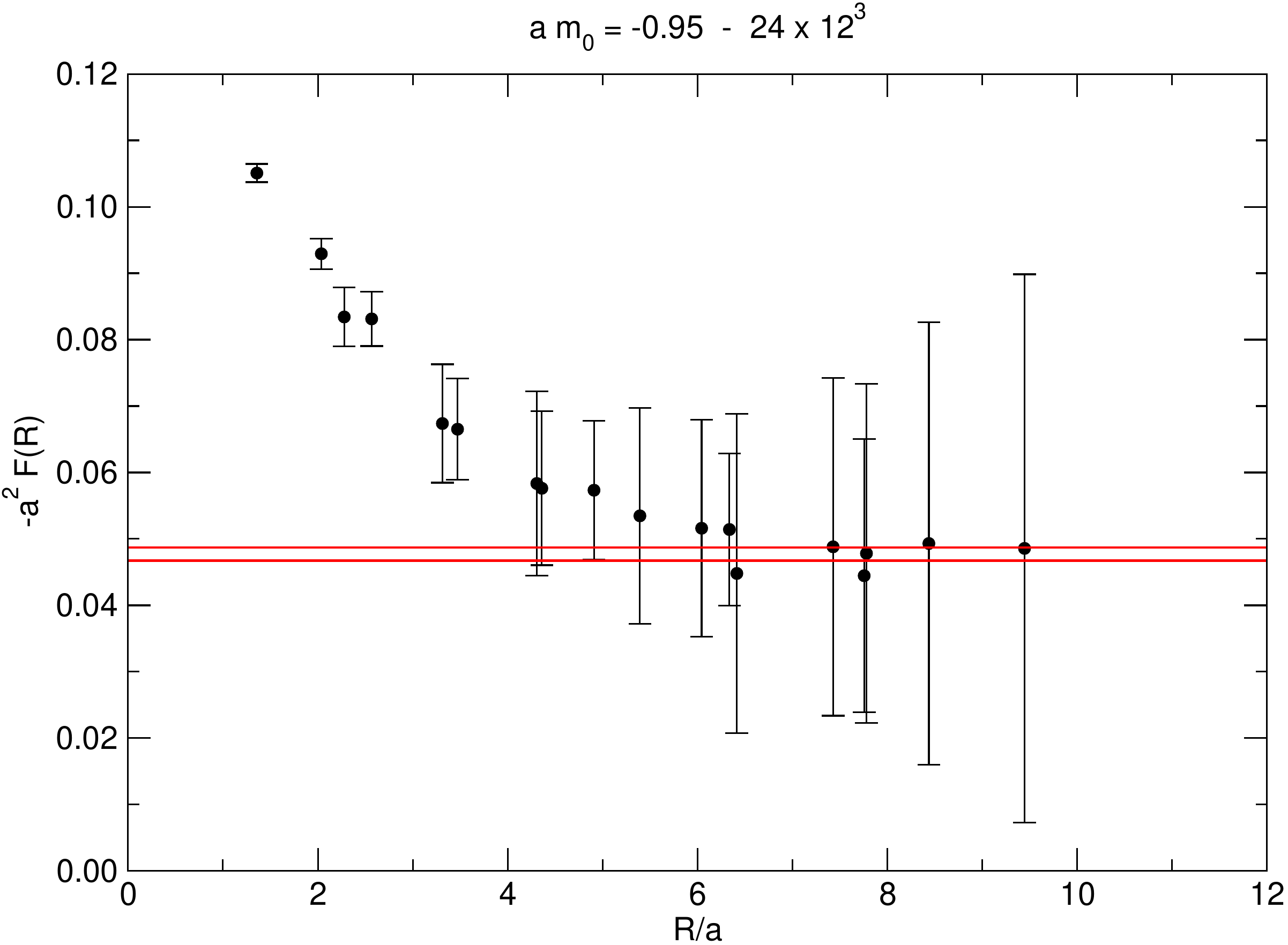}
}\parbox{.5\textwidth}{
\includegraphics*[width=.4\textwidth]{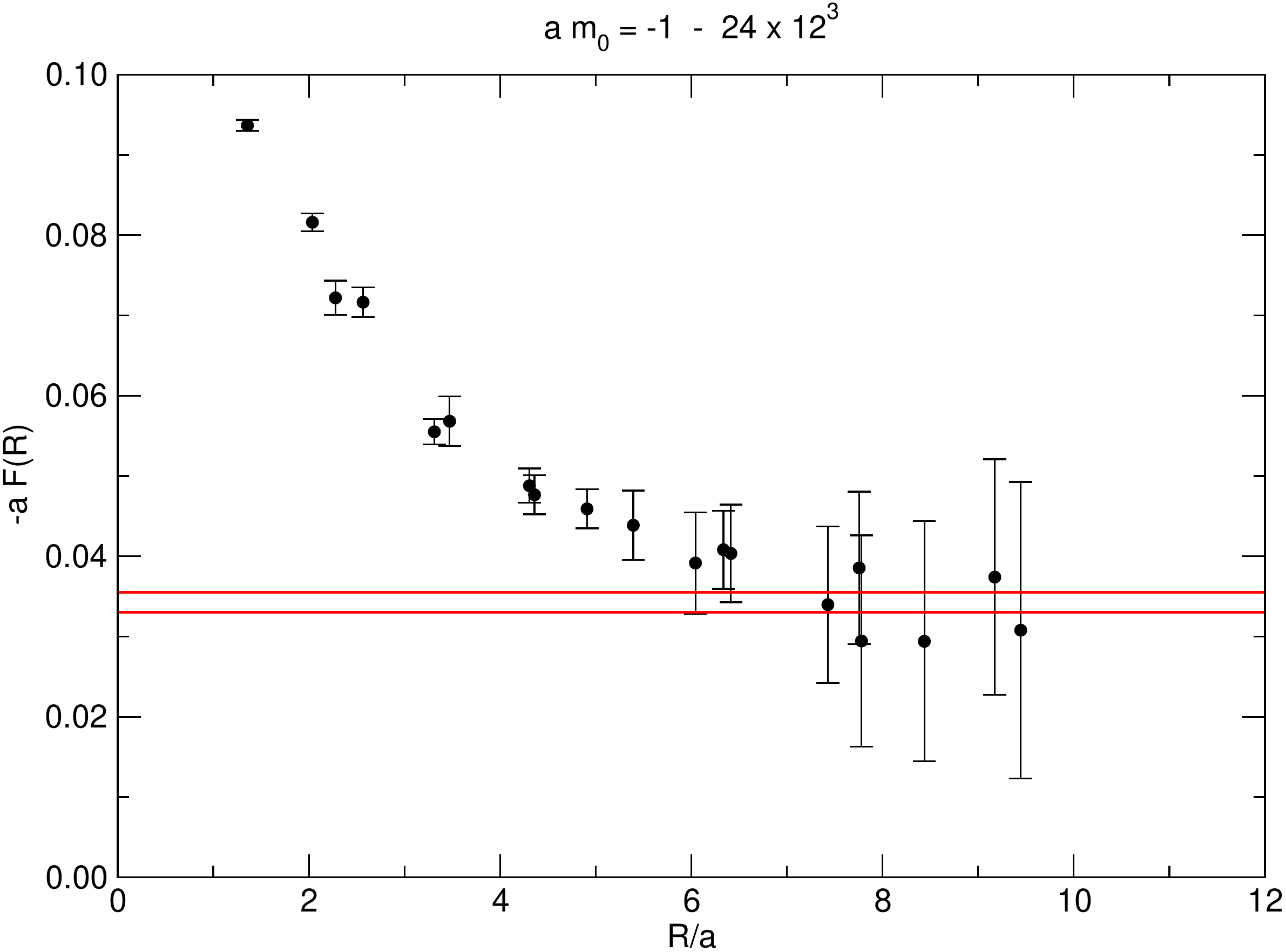}
}

\parbox{.5\textwidth}{
\includegraphics*[width=.4\textwidth]{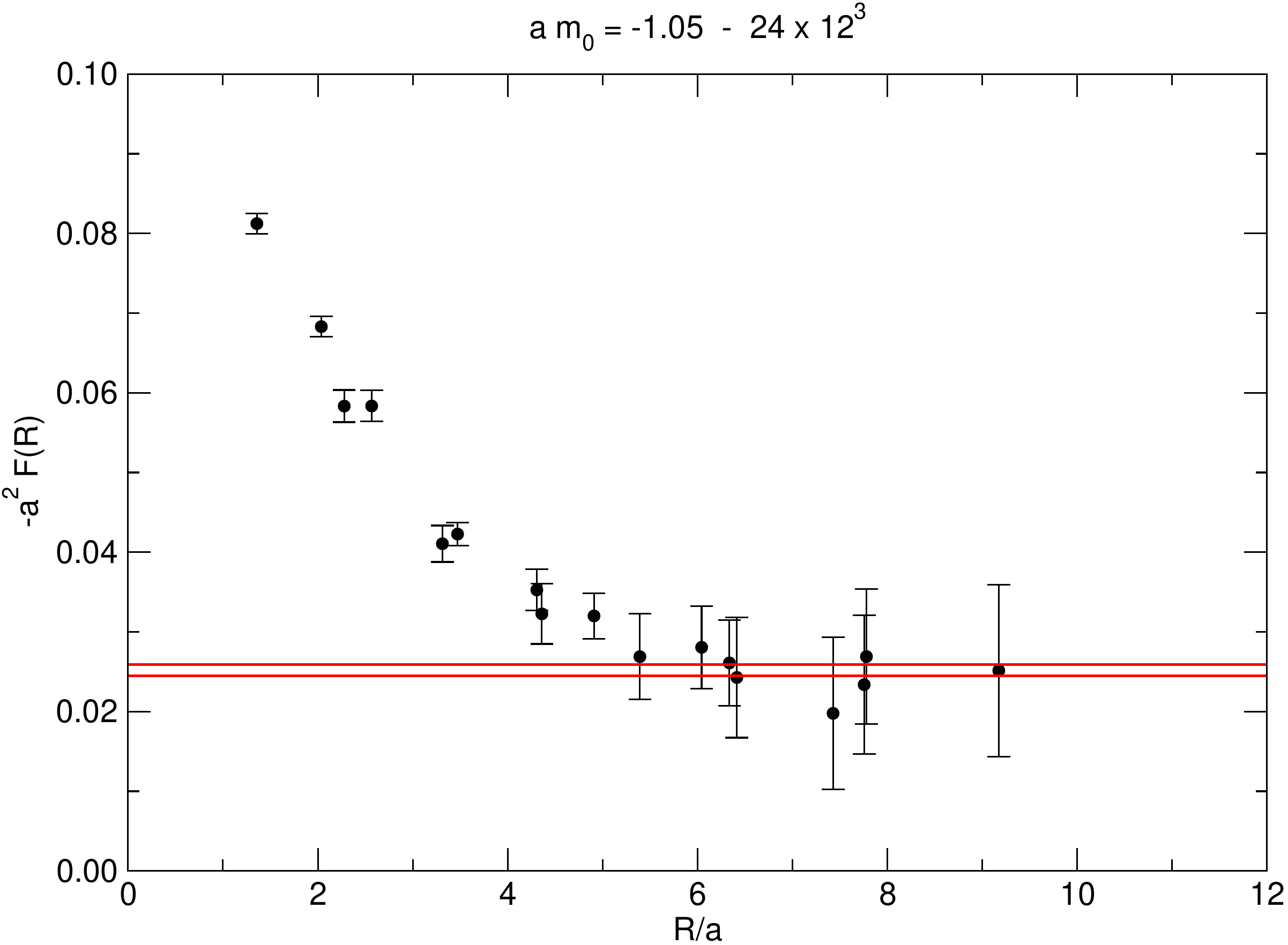}
}\parbox{.5\textwidth}{
\includegraphics*[width=.4\textwidth]{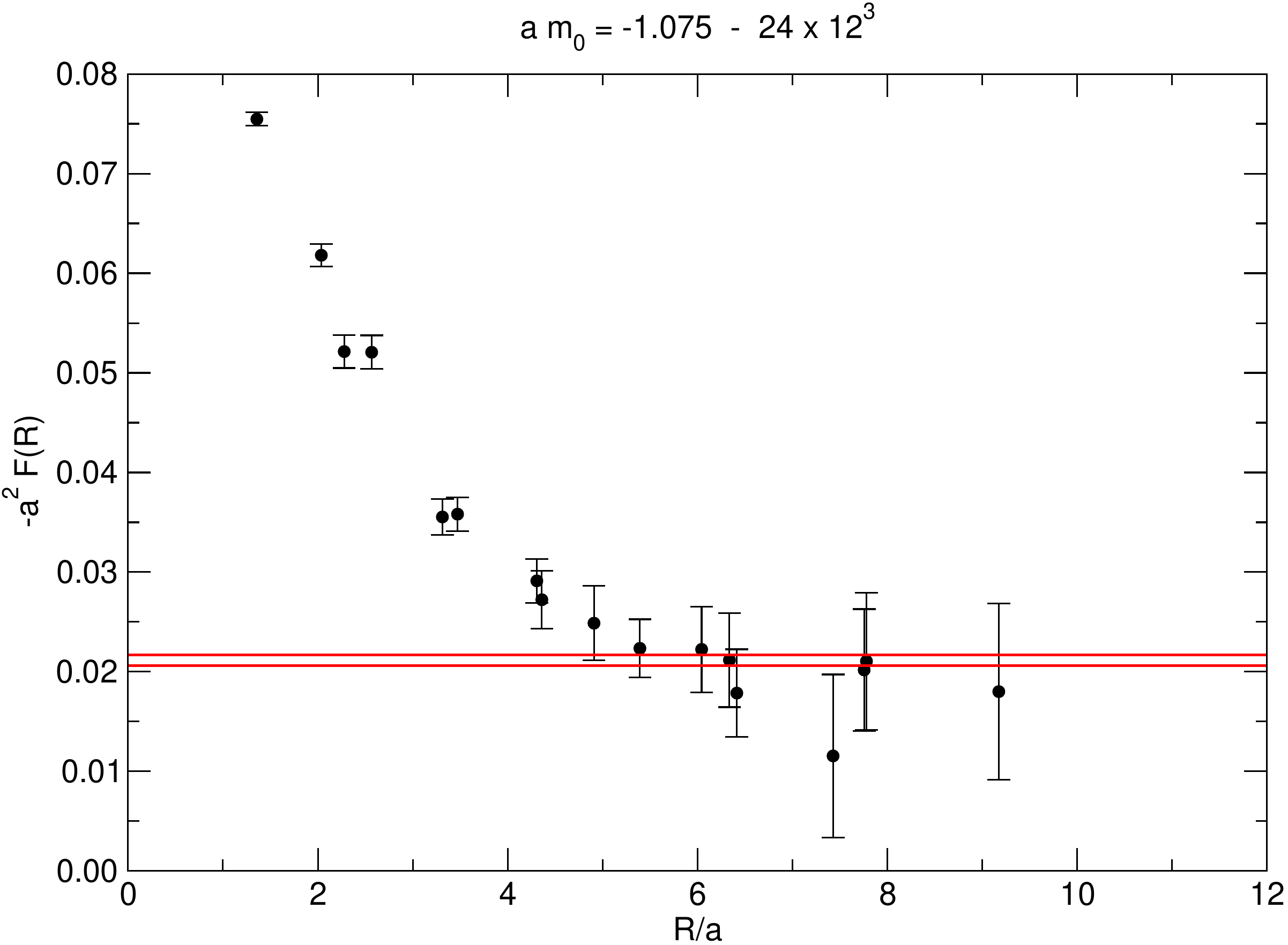}
}

\parbox{.5\textwidth}{
\includegraphics*[width=.4\textwidth]{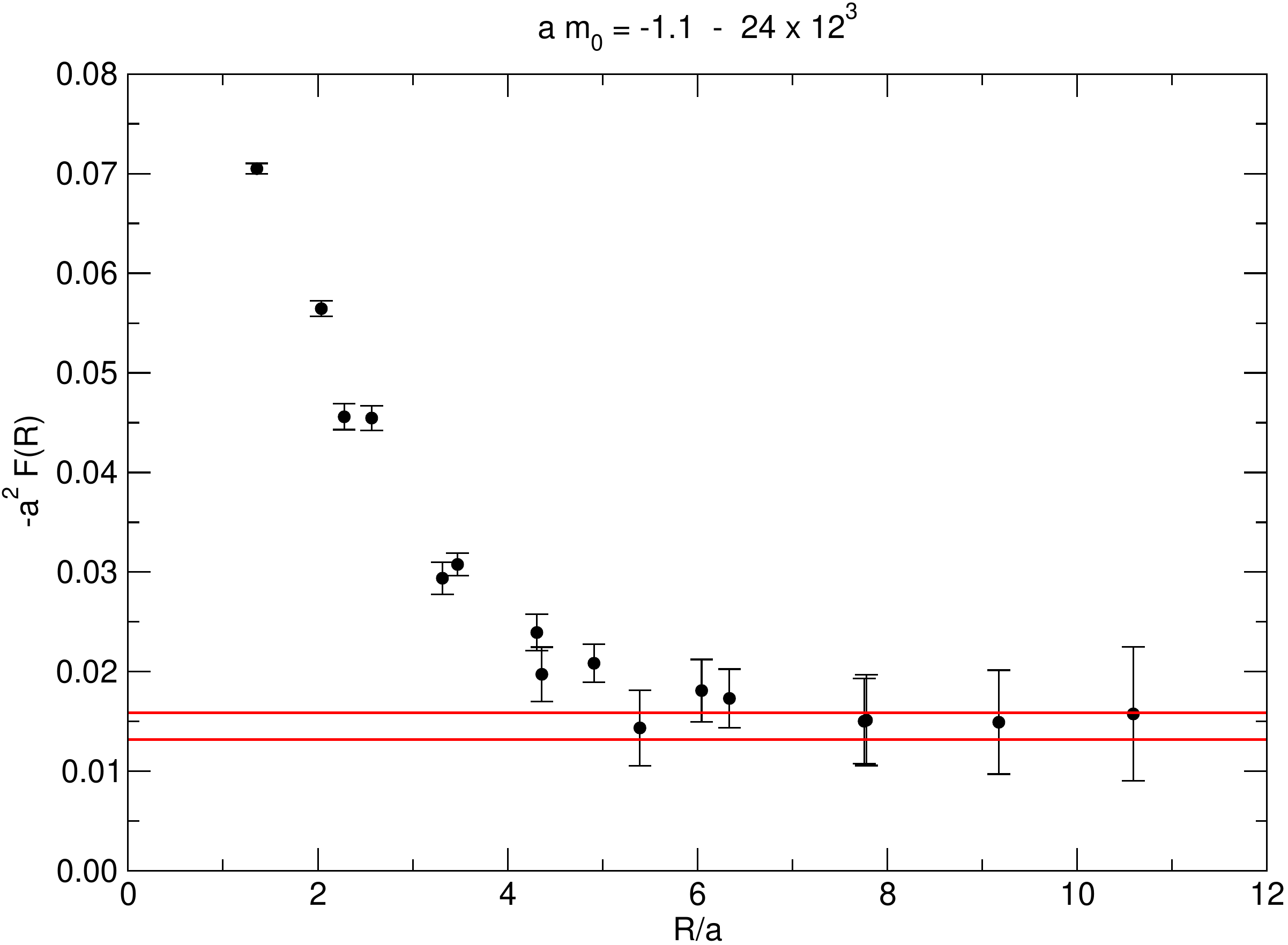}
}\parbox{.5\textwidth}{
\includegraphics*[width=.4\textwidth]{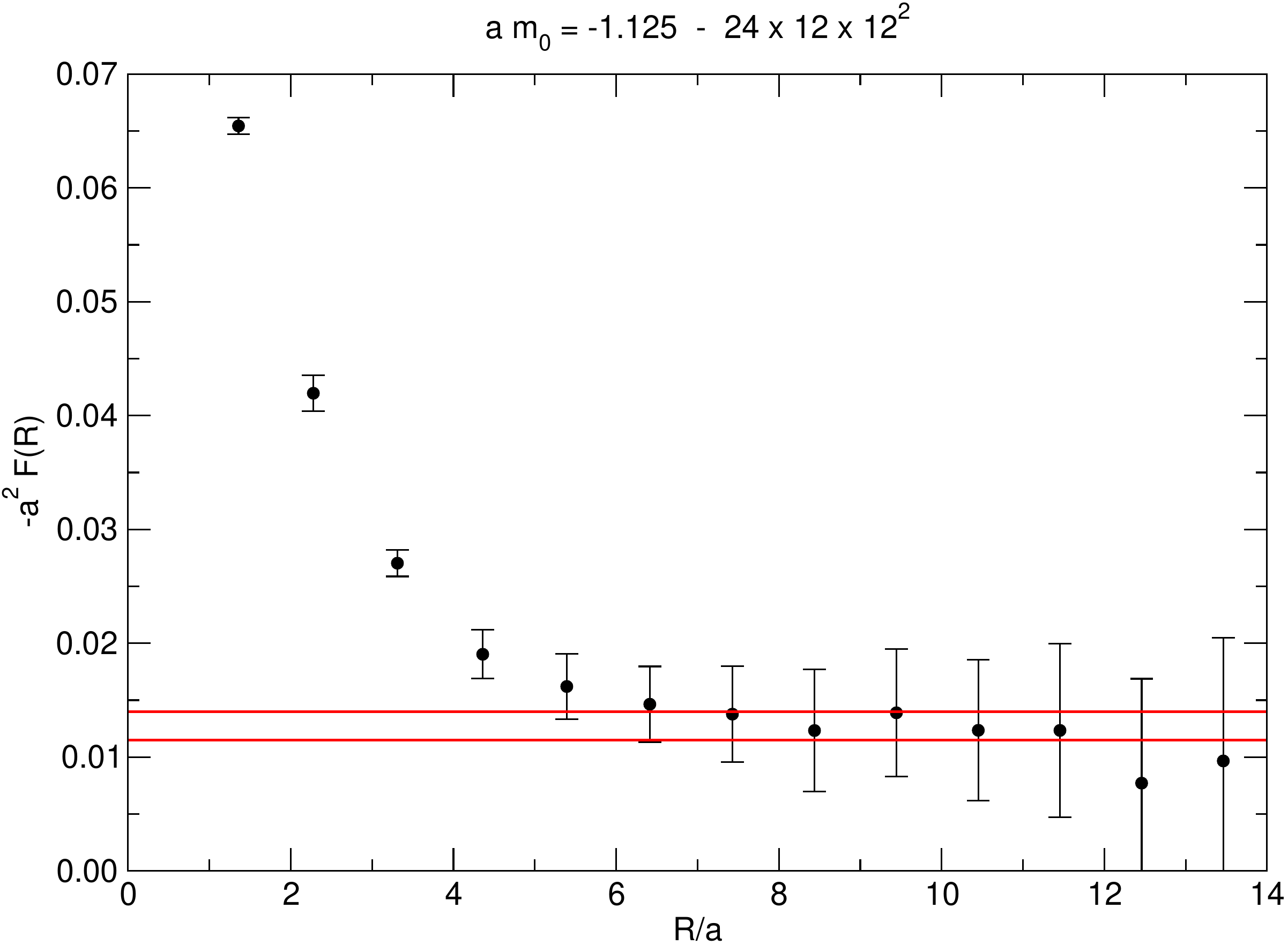}
}

\parbox{.5\textwidth}{
\includegraphics*[width=.4\textwidth]{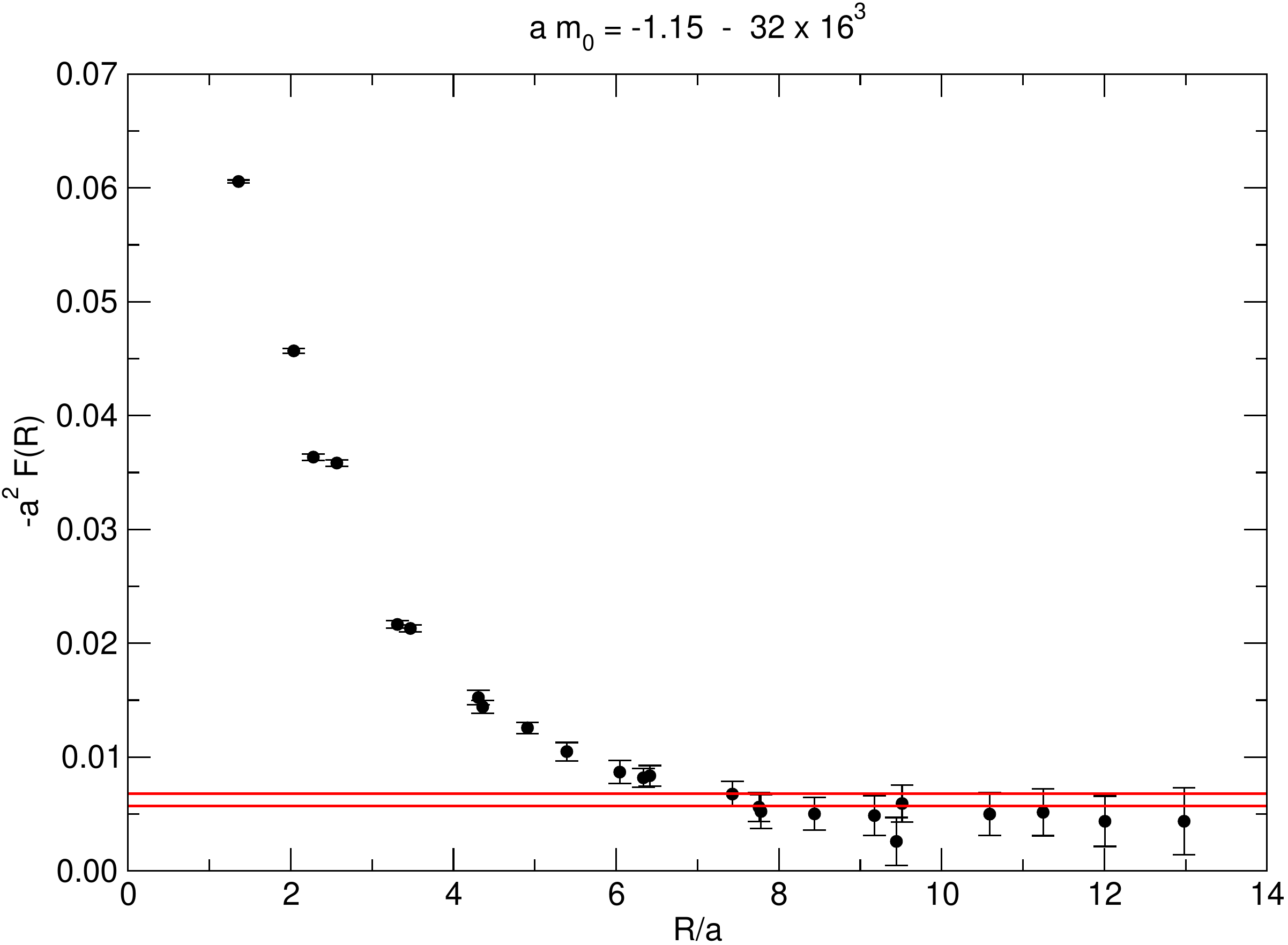}
}

\caption{Forces computed with the method \textbf{Force2} (see
  Sec.~\ref{sect:lattice:wilson}). Superimposed is the string tension
  from Table~\ref{tab:inf_sigma}.}
\label{fig:allforces}
\end{figure}

The static potentials for all the simulations listed in the last
column of Table~\ref{tab:inf_sigma} are plotted in
Fig.~\ref{fig:allpotentials}. The corresponding forces are separately
plotted in Fig.~\ref{fig:allforces}. Although in principle the static
potential or the force can be used to extract the string tension, in
practice a reliable result can not be obtained from those quantities,
the most likely explanation being either that our data are not
accurate enough or that the lattice sizes explored are too small for
the plateau to be free from systematic errors. A variational procedure
like the one used for extracting the string tension from correlators
of Polyakov loops (described in Sec.~\ref{sect:lattice:string}) might
be helpful also for the static potential computation. Although we are
unable to perform a comparison between the string tensions extracted
with the two different methods, we can still check that the static
potentials and the forces are compatible with the string tensions
reported in Table~\ref{tab:inf_sigma}.

We fit the static potential with the function:
\begin{equation}
V(R) = \sigma R + \mu + \frac{c}{R} \ ,
\end{equation}
assuming the string tensions shown in Table~\ref{tab:inf_sigma}, in
the range $R \ge 3a$. The results of the fits are shown in
Fig.~\ref{fig:allpotentials}. Since we do not want to assume at this
stage a particular effective string theory (and anyhow the HYP
smearing introduces spurious $1/R$ effects), the coefficient of the $1/R$ term
becomes an extra parameter in the fitting procedure. In all the cases
we have investigated, the string tension computed via Polyakov loop
correlators captures correctly the large distance behavior of the
static potential.

In Fig.~\ref{fig:allforces}, the forces are plotted together with the
values of the string tension. Although the errors on the force are in
some cases quite large and only qualitative statements are possible,
it can be seen also in this case that the string tension computed via
Polyakov loop correlators always captures the large distance behavior
of the force itself. The force always shows a plateaux at large
distances, with a central value often in striking agreement with the
string tension computed from Polyakov loop correlators. This might
indicate that our analysis overestimates the statistical errors.

\section{Glueballs}
\label{sect:lattice:glueballs}
Glueball masses are extracted from a variational procedure similar to
the one used for Polyakov loops and based on the same fuzzying
scheme. At the link level, we consider a collection of closed
elementary loops transforming according to the irreducible
representations of the symmetry group of the cube, to which rotational
symmetry is broken on the lattice~\cite{Berg:1982kp}. The variational
procedure is then built by replacing the original links with those
obtained after smearing and blocking. Then, in each channel a matrix
of connected correlators is constructed, whose eigenvectors with the
highest eigenvalues are almost the pure eigenstates of the Hamiltonian
with the lowest masses. We have implemented this technique using as
starting operators the plaquette and the length-six planar closed
contour to build the $A$, $E$ and $T$ irreducible representations of
the cubic group. The lowest-lying state in the $A$ channel corresponds
to the lightest $0^{++}$ glueball in the continuum limit, while both
the $E$ and the $T$ lightest states give the lightest $2^{++}$
glueball mass in the continuum limit.

\begin{figure}[ht]
\begin{minipage}[b]{0.45\linewidth}
\centering
\includegraphics*[width=0.9\textwidth]{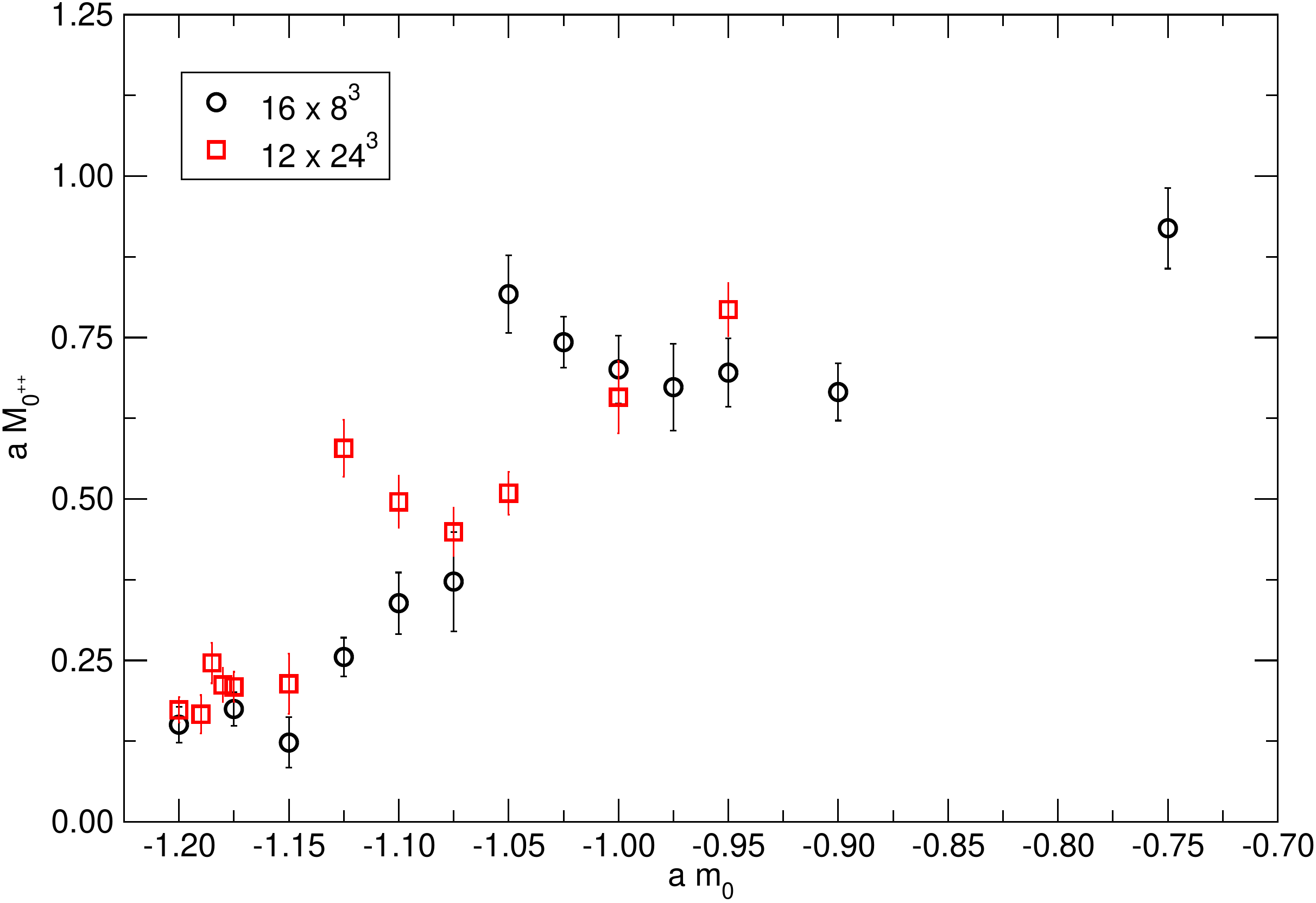}
\caption{The mass of the $0^{++}$ glueball in lattice units,
  $aM_{0^{++}}$, measured at various values of bare quark mass $am_0$ on
  a $16 \times 8^3$ and on a $24 \times 12^3$ lattice.}
\label{fig:mass_g0}
\end{minipage}
\hspace{0.05\textwidth}
\begin{minipage}[b]{0.45\linewidth}
\centering
\includegraphics*[width=0.9\textwidth]{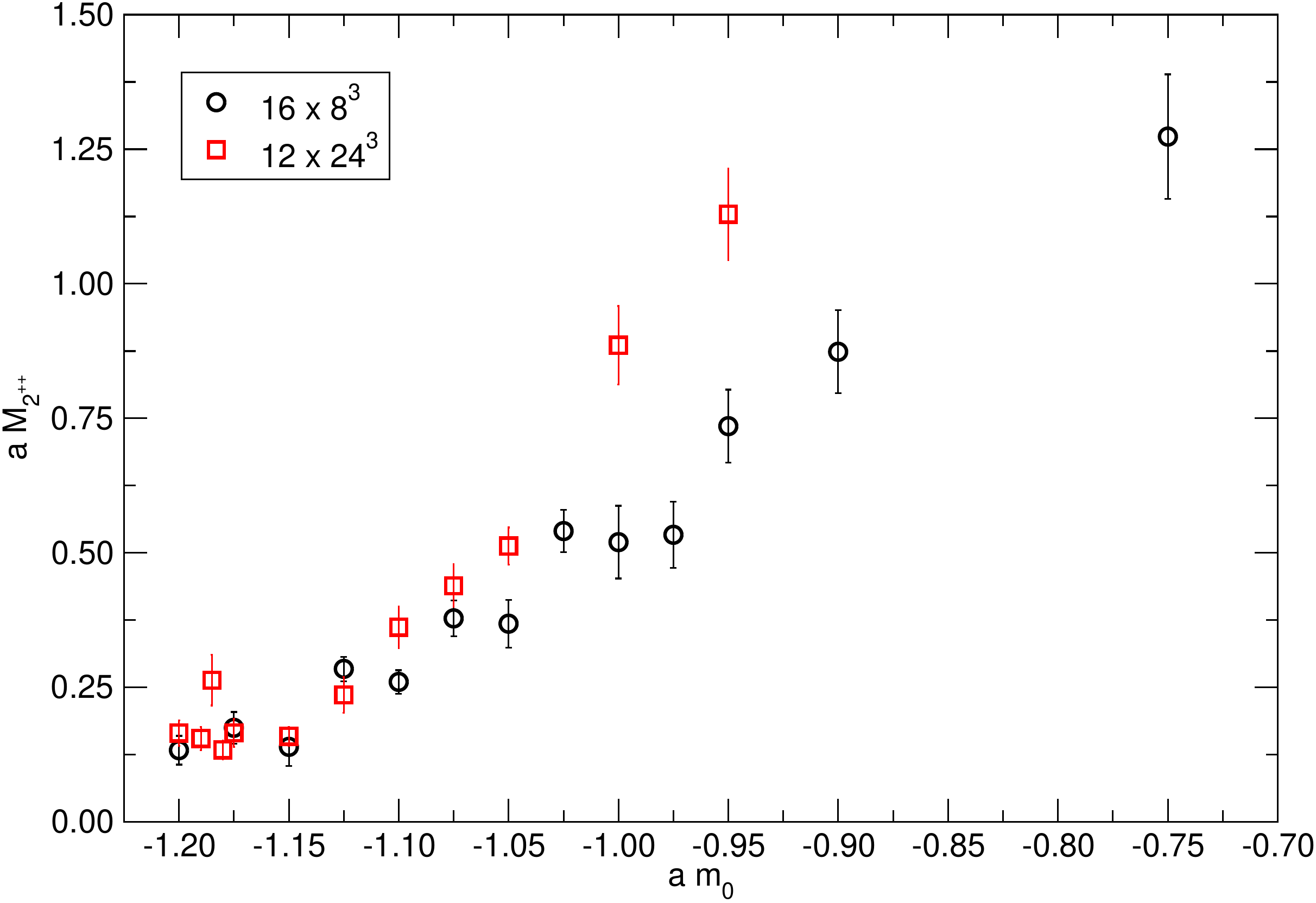}
\caption{The mass of the $2^{++}$ glueball in lattice units,
  $aM_{2^{++}}$, measured at various values of bare quark mass $am_0$ on
  a $16 \times 8^3$ and on a $24 \times 12^3$ lattice.}
\label{fig:mass_g2}
\end{minipage}
\end{figure}
Our results for the $0^{++}$ glueball are illustrated in
Fig.~\ref{fig:mass_g0}. We discuss only the results on the $16 \times
8^3$ and the $24 \times 12^3$ lattices, since on the $32 \times 16^3$
lattice for all values of the simulated bare fermion masses our system
is in the A-phase. As it can be seen from the plot, as $am_0$ is
decreased, in the S-phase $aM_{0^{++}}$ monotonically decreases. When
the system starts to develop double peaks for the Polyakov loop
distribution in a spatial direction, the mass of the $0^{++}$ glueball
first goes to a short plateau and then moderately increases, to drop
dramatically to much lower values at lower $am_0$. The overlap
between the masses measured on the two lattices at $am_0 = -0.95$
suggests that as long as we stay in the symmetric phase, finite size
effects are under control. Hence, as our best estimate for the
infinite volume limit of $aM_{0^{++}}$, we take the values on the $16
\times 8^3$ lattice for $am_0 > -0.95$ and the values on the $24 \times
12^3$ lattice for $-1.05 \le am_0 \le -0.95$. Since for $am_0 < -1.05$
in both cases our system is in the A-phase, we do not take into
account the corresponding values of $am_{0++}$ in the following
analysis.

The mass of the $2^{++}$ glueball as a function of $am_0$ is shown in
Fig.~\ref{fig:mass_g2}. These results have been obtained using
operators transforming according to the $E$ representation of the cubic
group. Our analysis in the $T$ channel gives compatible results in all
cases. Our data show that the $2^{++}$ glueball is heavier than the
$0^{++}$ in the symmetric phase, but dramatically decreases to very
low values of the mass (well below the mass of the $0^{++}$) at the
onset of the A-phase.  Deeper in the A-phase, the two states
appear to be degenerate. As the figure shows, no good control over
finite size effects can be reached on our lattices for the $2^{++}$
mass. For the sake of completeness, we still provide an estimate for
its mass at infinite volume, but this is likely to be quite
rough. Hence, the $2^{++}$ glueball will play a marginal role in the
interpretation of our results.

Our numerical estimates of $aM_{0^{++}}$ and $aM_{2^{++}}$ in the infinite
volume limit are
reported in Tab.~\ref{tab:inf_glueballs}. The degeneracy between the two
states at $m = -1.05$ together with the impossibility of establishing whether
the system is in the S-phase (see Tab.~\ref{tab:poly:12}) would suggest to
disregard glueball masses at this value of the bare mass. However, since this
point was part of our analysis in Ref.~\cite{DelDebbio:2009fd}, where the lower
statistics masked the issue, in order to facilitate a comparison with our
previous work, we chose to keep it also in our current analysis. The reader
should bear this in mind for the discussion of our results.

\begin{table}[ht]
\centering
\begin{tabular}{c|c|c}
\hline
$-am_0$ & $aM_{0^{++}}$ & $aM_{2^{++}}$\\
\hline
\hline
- 0.25  & 1.159(98)  & 2.18(22)  \\
0.25    & 1.108(97)  & 1.92(22)  \\
0.5     & 1.045(70)  & 1.93(19)  \\
0.75    & 0.919(63)  & 1.27(12)  \\
0.9     & 0.666(44)  & 0.874(77) \\
0.95    & 0.793(41)  & 1.129(85) \\
1       & 0.658(56)  & 0.886(73) \\
1.05    & 0.510(33)  & 0.513(35) \\
\hline
\end{tabular}
\caption{Infinite volume estimates of $aM_{0^{++}}$ and $aM_{2^{++}}$. Values extracted on a $16 \times 8^3$ lattice have been used for $am_0 > -0.95$ and values extracted on a $24 \times 12^3$ lattice for $am_0 \le -0.95$.}
\label{tab:inf_glueballs}
\end{table}

\begin{figure}[ht]
\centering
\includegraphics*[width=.6\textwidth]{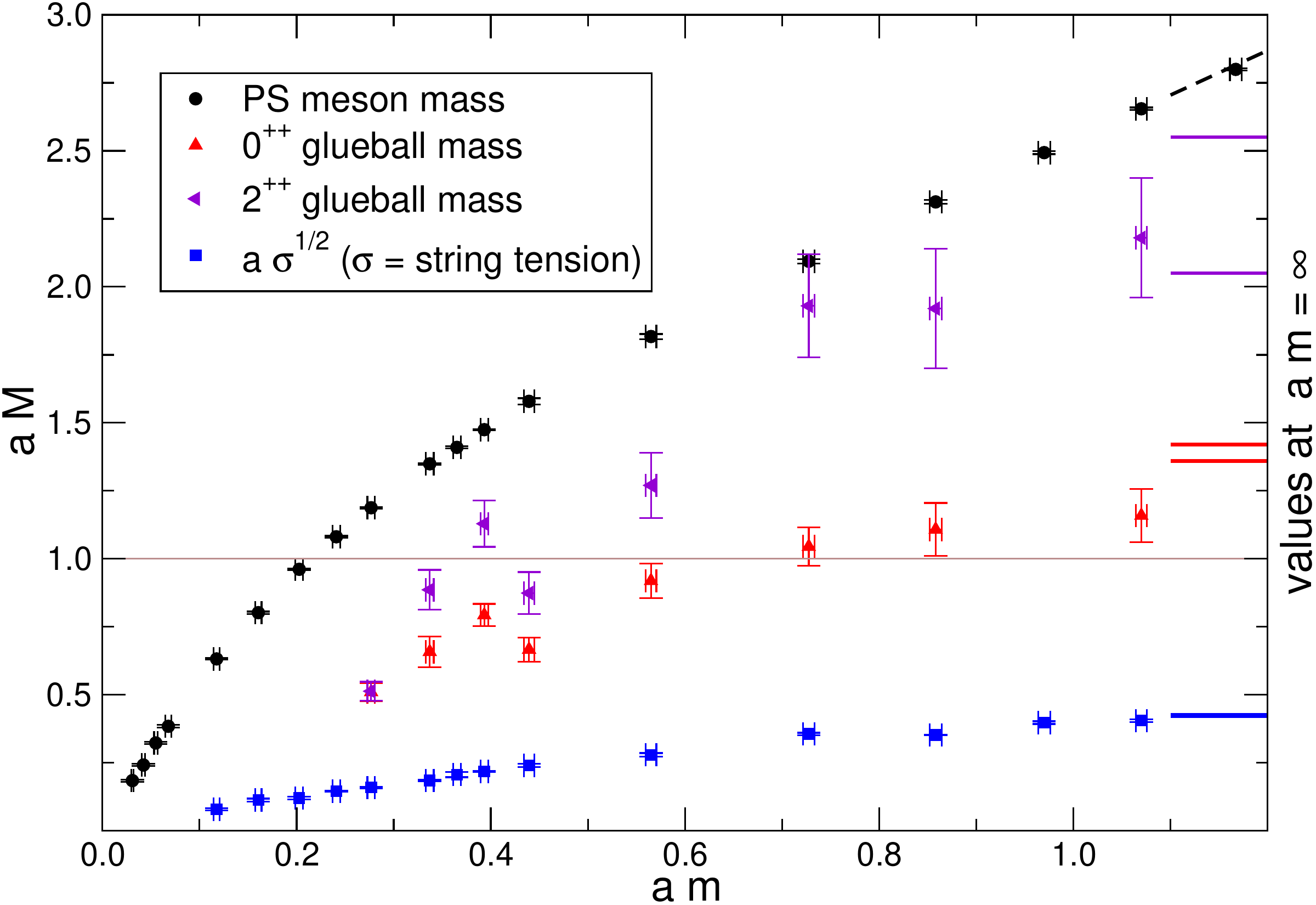}
\caption{The spectrum of the theory as a function of the PCAC mass
  $am$. The mass of the vector is not shown, since on the scale of the
  figure this state appears to be degenerate with the PS.}
\label{fig:spectrum_all}
\end{figure}

\section{Hyperscaling scenario and locking scale}
\label{sect:locking}
Our infinite volume estimates for the PS (at each value of the bare mass,
we choose the PS mass computed on the largest volume in \cite{noi}), the
$0^{++}$ and the $2^{++}$ glueball masses and $\sigma^{1/2}$ as a function
of the PCAC mass $am$ (see Refs.~\cite{DelDebbio:2008zf,noi} for a
definition of this quantity) are reported in Fig.~\ref{fig:spectrum_all}.
As noticed in Ref.~\cite{DelDebbio:2009fd}, the data show a clear
hierarchy in the spectrum, with the mesonic scale well above the
gluonic scale. Since over the range of investigated masses
$a\sigma^{1/2}$ changes by a factor of five, the effect of the fermion
determinant as the mass is decreased is an essential component of the
dynamics in this theory. Hence, the simple quenched scenario,
according to which the theory would be QCD-like and the hierarchy in
the spectrum is due to large fermion masses, can be excluded. In fact,
the spectrum looks similar to the hyperscaling scenario at high locking mass
$M_\mathrm{lock}$ sketched in Fig.~\ref{fig:locking} (right). In this
section we shall show that indeed that scenario provides the right
description of the spectrum of this theory.

\begin{figure}[ht]
\centering
\includegraphics*[width=.6\textwidth]{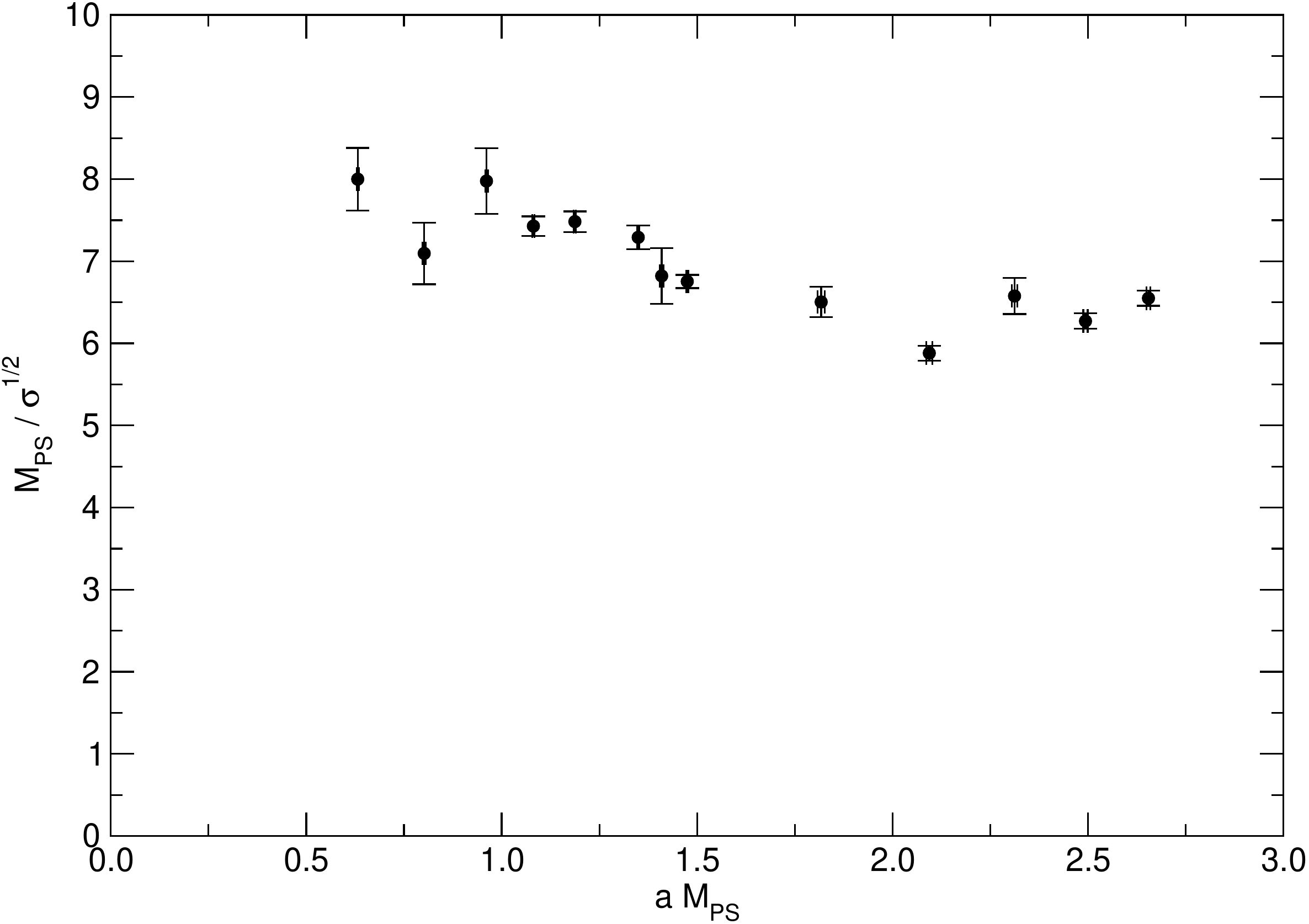}
\caption{The ratio $M_\mathrm{PS}/\sigma^{1/2}$ as a function of
  $M_\mathrm{PS}$.}
\label{fig:mps_over_sqrtsigma}
\end{figure}

\begin{figure}[ht]
\centering
\includegraphics*[width=.6\textwidth]{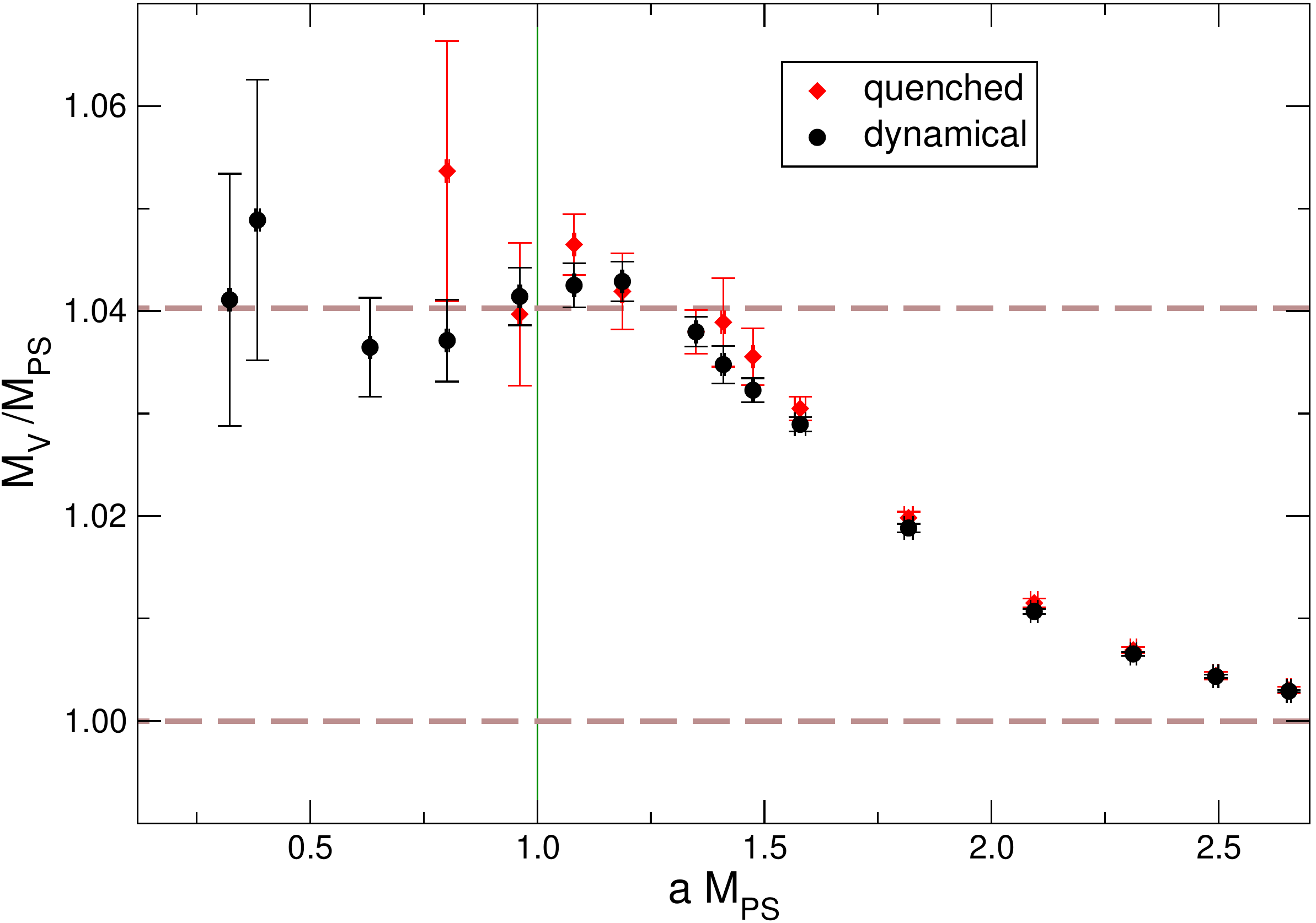}
\caption{Comparison of the ratio $M_\mathrm{V}/M_\mathrm{PS}$ as
  function of $aM_\mathrm{PS}$ in the quenched and the dynamical
  theory. The dynamical ratio approaches the infinite-mass value of
  one at large $aM_\mathrm{PS}$ and develops a plateau (signaling
  hyperscaling) at $aM_\mathrm{PS} \le 1.25$.}
\label{fig:mv/mrho_comp}
\end{figure}

Up to subleading corrections, the hyperscaling scenario implies the
independence of ratios of physical quantities from the fermion mass
in the scaling region. In Fig.~\ref{fig:mps_over_sqrtsigma} we plot
the ratio $M_\mathrm{PS}/\sigma^{1/2}$. This quantity shows a plateau
$M_\mathrm{PS}/\sigma^{1/2} \simeq 7.5$ for $a M_\mathrm{PS} \le
1.25$, supporting the idea that gluonic and fermionic masses are not
parametrically independent in this region but are both proportional
to the RG-invariant fermion mass $M$ (again, this is what we are
calling locking). The independence from $M$ of the ratio of spectral
quantities in the scaling region can be observed also in the ratio
$M_\mathrm{V}/M_\mathrm{PS}$ (Fig.~\ref{fig:mv/mrho_comp}). Again a
plateau develops for $a M_\mathrm{PS} \le 1.25$. The existence of these
plateaux is a clear indication of the spectrum behaving as predicted
by hyperscaling. Moreover, the value of both
$M_\mathrm{PS}/\sigma^{1/2}$ and $M_\mathrm{V}/M_\mathrm{PS}$ in the
scaling region suggests that the locking scale $M_\mathrm{lock}$ is
large. If this is the case, SU(2) gauge theory with two adjoint Dirac
fermions should look like an heavy fermion system for all values of
the fermion mass.

In order to verify this scenario, we can compare our dynamical results
with results obtained in the quenched theory. In this process, a
crucial point is to match properly the bare parameters in the two
theories, since the physics of the dynamical and quenched systems is
expected to be equivalent once the physical scale is matched (see
Eq.~\eqref{eq:locking}). For this reason, we need
to tune the bare parameters of the quenched simulations, namely the
gauge coupling $\beta^{(q)}$, and the mass of the valence fermion
$am^{(q)}_0$, so that we can match two independent quantities between
the two theories. For this matching, it is convenient to choose two
observables with a sharp dependency on each of the bare parameters, in
order to achieve the best possible tuning between the two theories. In
our study, we have required the quantities $a\sigma^{1/2}$ and
$aM_\mathrm{PS}$ to be equal in the dynamical and in the quenched
systems. Given that in the dynamical theory with adjoint fermions the
fundamental string cannot break, this quantity is a natural candidate
to fix the coupling $\beta^{(q)}$ in the quenched simulation. As far
as the valence quark mass $a m_0^{(q)}$ is concerned, we chose to
determine the quenched value by fixing the pseudoscalar mass, because
in the fermionic sector this is the quantity we have the best control
upon.

The procedure to compare the quenched and dynamical
theories requires the following steps:
\begin{enumerate}
\item find the value $\beta^{(q)}$ of the coupling for the quenched
  theory, in such a way that the string tension in lattice units
  matches the dynamical value;
\item find the value $am_0^{(q)}$ of the bare fermion mass for the
  quenched theory, in such a way that the PS meson mass matches the
  dynamical value $aM_\mathrm{PS}$;
\item compute the rest of the spectrum, for example the ratio of the
  PS and V meson masses or the glueball masses, in the dynamical
  theory with bare parameters $(\beta,am_0)$ and in the quenched theory
  with bare parameters $(\beta^{(q)},am_0^{(q)})$, and compare the
  results.
\end{enumerate}

In practice the program as outlined above requires a fine tuning of
the bare parameters, and turns out to be a highly expensive
computational task. Instead of an exact matching of the parameters, we
have performed a scan in the parameters of the quenched theory. The
lattice parameters at which quenched simulations have been performed
are reported in Tab.~\ref{tab:quenched}. The range of the scan of
$\beta^{(q)}$ is chosen in order to include all the string tension
values of our dynamical simulations. The upper
bound of this window is simply given by $\beta\equiv2.25$, the value
of $\beta$ for our dynamical simulations. This is a consequence of the
string tension being an increasing function of the bare fermion mass,
so an infinite mass simulation corresponds to a string tension of a
pure gauge system at the same $\beta\equiv2.25$.

For each choice of $\beta^{(q)}$ we measure the string tension and the
$0^{++}$ and $2^{++}$ glueball masses (following
e.g. Ref.~\cite{Lucini:2001nv}). Quenched glueball masses have been
interpolated using the ansatz
\begin{equation} 
\frac{M_\mathrm{G}}{\sigma^{1/2}}  = A_0 + A_1 a^2\sigma \ ,
\end{equation}
with $A_0$ and $A_1$ respectively the leading (constant) and
subleading (${\cal O}(a^2)$) coefficients in the extrapolation to the
continuum limit.

On the same gauge configurations, we measure the quenched PS mass and
the $M_\mathrm{V}/M_\mathrm{PS}$ ratio, for a set of values of
$am^{(q)}_0$ covering the entire interval of PS masses appearing in
the dynamical calculation. We then create an interpolating function
for the central value of the ratio $M_\mathrm{V}/M_\mathrm{PS}$. 

To obtain an error on this estimate, we create two other interpolating
functions for the maximal and the minimal value of the quenched
estimate $M_\mathrm{V}/M_\mathrm{PS}$ set by the statistical error, so
that for each choice of the pair $(M_\mathrm{PS},\sigma)$, we can read
the corresponding range of values for $M_\mathrm{V}/M_\mathrm{PS}$. To
take into account the indetermination in our estimate of $a M_\mathrm{PS}$ and
$a^2 \sigma$, we consider a region within one sigma around the central
value for those quantities: in this region the difference between the
maximum value of the maximal interpolating function and the minimum of
the minimal interpolating function provides us with an estimate for
the error on $M_\mathrm{V}/M_\mathrm{PS}$ in the quenched theory.

By means of the interpolating functions, we can read the value of
$M_\mathrm{G}/\sqrt{\sigma}$ and $M_\mathrm{V}/M_\mathrm{PS}$ in the
quenched simulations at values of $a \sqrt{\sigma}$ and $a M_\mathrm{PS}$
(the latter being relevant only for the $M_\mathrm{V}/M_\mathrm{PS}$
ratio) obtained in the dynamical simulations.

\begin{figure}[ht]
\centering
\includegraphics*[width=.6\textwidth]{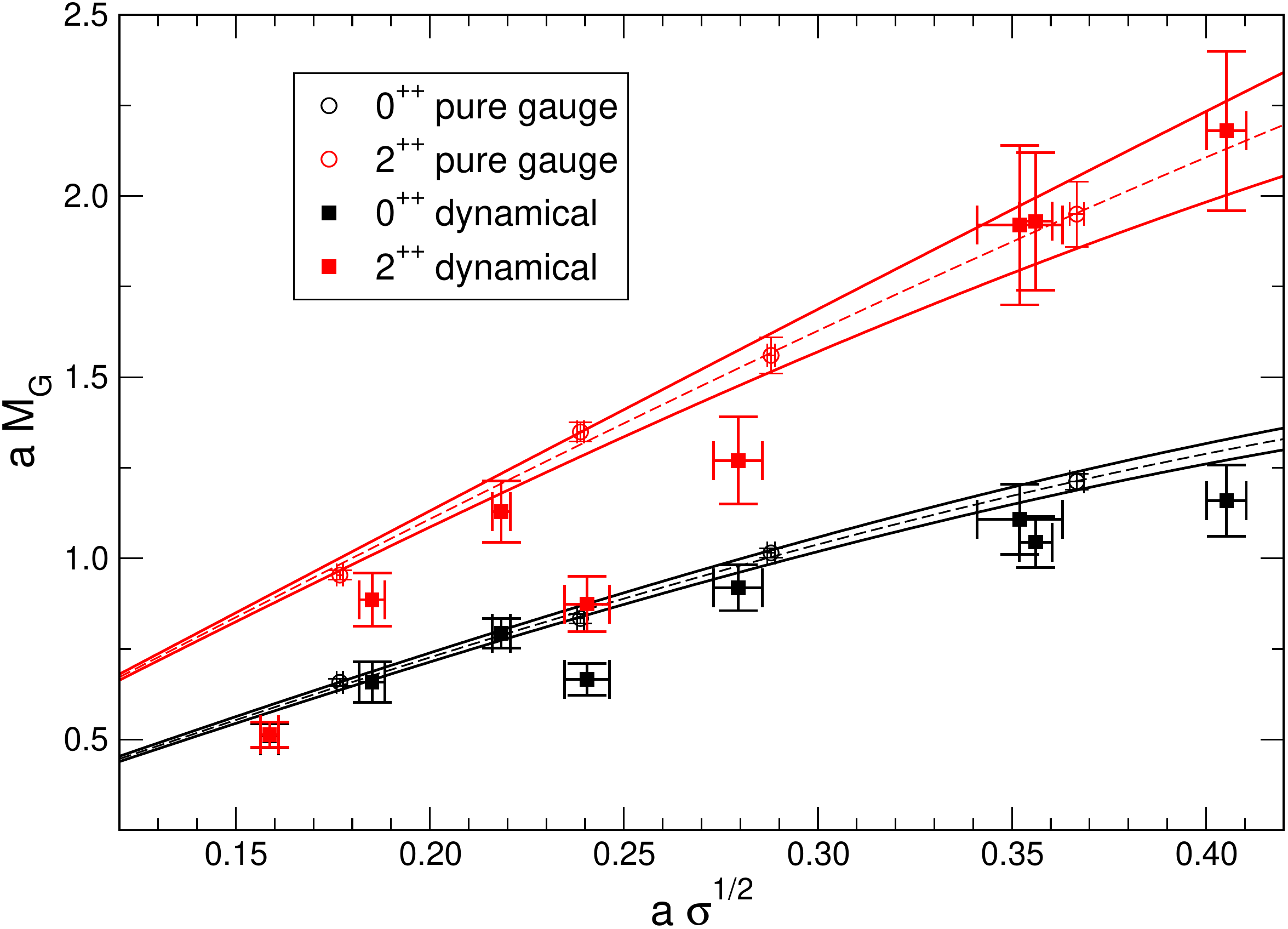}
\caption{Comparison of the masses of glueballs of the dynamical theory
  with the interpolating functions of the quenched theory. Among the plotted data, the dynamical points at $a\sigma^{1/2}=0.2405(58)$ and $a\sigma^{1/2}=0.1587(23)$ correspond to the lightest masses respectively on the $16 \times 8^3$ and $24 \times 12^3$ lattices. Hence, the glueball masses for those string tensions are affected by the largest finite-size effects.}
\label{fig:glue_comp}
\end{figure}

In Tab.~\ref{tab:glue_quenc} we report the values of the mass of the
glueballs for the quenched theory at the values of the string tension
obtained in the dynamical theory. In Fig.~\ref{fig:glue_comp} we show
the comparison between the dynamical glueball values and the
interpolating functions obtained from the quenched theory. Except for
the last point, for which, as discussed in
Sect.~\ref{sect:lattice:glueballs}, the dynamical simulations are
probably in the A-phase, the agreement between the quenched and
dynamical spectra at the same physical scale (in units of the
ultraviolet cutoff) is striking. This supports the idea that the
low-energy dynamics of the theory with dynamical fermions is well
described by a pure Yang-Mills theory, which is evidence for a locking
mechanism with a large $M_\mathrm{lock}$ taking place in this theory.

The relevant interpolated quenched results for the ratio
$M_\mathrm{V}/M_\mathrm{PS}$ as a function of $M_\mathrm{PS}$ are
reported in Tab.~\ref{tab:mv/mrho_quenc} and compared with the
dynamical results in Fig.~\ref{fig:mv/mrho_comp}. The quenched and
dynamical data have a remarkable overlap for all the points except for
the last one, where strong finite-size effects are expected to affect
both the quenched and the dynamical simulations. Together with the
plateau in the ratio developing for $aM_\mathrm{PS} \le 1.25$, the comparison
confirms once again a locking mechanism taking place at large
$M_\mathrm{lock}$.

\begin{table}[ht]
\centering
\begin{tabular}{cccc}
\hline
$\beta^{(q)}$ & $a\sigma^{1/2}$ & V & $am_0^{(q)}$ \\
\hline
\hline
\multirow{3}{*}{2.25} & \multirow{3}{*}{0.4231(25)} &\multirow{3}{*}{$24\times12^3$}& -1.65,-1.6,-1.55,-1.5,-1.45,-1.4,-1.35,-1.3, \\
& & & -1.25,-1.2,-1.175,-1.15,-1.125,-1.1,-1.075,\\
& & & -1.05,-1.025,-1.0,-0.75,-0.5,-0.25,0.0,0.25,0.5 \\
\hline
\multirow{2}{*}{2.4265} & \multirow{2}{*}{0.2388(9)} &\multirow{2}{*}{$24\times12^3$}& -1.2,-1.175,-1.15,-1.125,-1.1,-1.075,-1.05, \\
& & & -1.025,-1.0,-0.75,-0.5,-0.25,0.0,0.25,0.5 \\
\hline
\multirow{2}{*}{2.5115} & \multirow{2}{*}{0.1768(8)} &\multirow{2}{*}{$24\times12^3$}& -1.2,-1.175,-1.15,-1.125,-1.1,-1.075,-1.05, \\
& & & -1.025,-1.0,-0.75,-0.5,-0.25,0.0,0.25,0.5 \\
\hline
\multirow{2}{*}{2.6} & \multirow{2}{*}{0.13395(62)} &\multirow{2}{*}{$24\times12^3$}&  -1.2,-1.175,-1.15,-1.125,-1.1,-1.075,-1.05, \\
& & & -1.025,-1.0,-0.75,-0.5,-0.25,0.0,0.25,0.5 \\
\hline
\multirow{2}{*}{2.62} & \multirow{2}{*}{0.1258(7)} &\multirow{2}{*}{$24\times12^3$}&   -1.3,-1.25,-1.2,-1.15,-1.1,-1.05,-1.0,-0.75, \\
& & & -0.5,-0.25,0.0,0.25,0.5 \\
\hline
\multirow{2}{*}{2.68} & \multirow{2}{*}{0.1035(7)} &\multirow{2}{*}{$32\times32^3$}&  -1.25,-1.2,-1.15,-1.1,-1.05,-1,-0.95,-0.9,\\
& & & -0.85,-0.8,-0.75,-0.5 \\
\hline
\end{tabular}
\caption{Bare parameters and volumes used for quenched simulations. For each $\beta^{(q)}$ we report also the measured string tension.}
\label{tab:quenched}
\end{table}

\begin{table}[ht]
\centering
\begin{tabular}{c|c|c}
\hline
$a\sigma^{1/2}$ &  $aM_{0^{++}}^{(q)}$ & $aM_{2^{++}}^{(q)}$ \\
\hline
\hline
0.4053(51) & 1.30(4)  & 2.13(15)   \\
0.352(11)  & 1.18(5)  & 1.89(14)   \\
0.3561(42) & 1.185(35) & 1.90(11)  \\
0.2794(63) & 0.975(35) & 1.53(8)   \\
0.2405(58) & 0.855(35) & 1.325(65) \\
0.2184(23) & 0.785(15) & 1.21(4)   \\
0.1851(33) & 0.675(25) & 1.025(35) \\
0.1587(23) & 0.585(15) & 0.885(25) \\
\hline
\end{tabular}
\caption{The results of the interpolation procedure for the quenched data of the mass of the glueballs at the values of $a\sigma^{1/2}$ equal to the ones found in the dynamical simulations. The errors on the third column have been obtained from the variation of the numerical results on the maximal and the minimal interpolating function.}
\label{tab:glue_quenc}
\end{table}

\begin{table}[ht]
\centering
\begin{tabular}{c|c|c}
\hline
$a\sigma^{1/2}$ & $aM_\mathrm{PS}$ & $M_\mathrm{V}^{(q)}/M_\mathrm{PS}^{(q)}$ \\
\hline
\hline
0.4053(51) & 2.6546(47)  & 1.00303(32)\\
0.3976(51) & 2.4936(57)  & 1.00442(38)\\
0.352(11)  & 2.3120(68)  & 1.00693(30)\\
0.3561(42) & 2.0939(80)  & 1.01152(44)\\
0.2794(63) & 1.8172(95)  & 1.01983(59)\\
0.2405(58) & 1.579(12)   & 1.0304(11) \\
0.2184(23) & 1.4748(24)  & 1.0355(27) \\
0.2066(97) & 1.4094(42)  & 1.0389(40) \\
0.1851(33) & 1.3493(28)  & 1.0379(21) \\
0.1587(23) & 1.1874(28)  & 1.0419(37) \\
0.1455(19) & 1.0811(31)  & 1.0464(30) \\
0.1205(56) & 0.9613(35) & 1.0396(70) \\
0.1130(54) & 0.8017(41) & 1.053(13)  \\
\hline
\end{tabular}
\caption{The results of the interpolation procedure for the quenched data of  $M_\mathrm{V}^{(q)}/M_\mathrm{PS}^{(q)}$   at the values of $a\sigma^{1/2}$ and $aM_\mathrm{PS}$ equal to the ones found in the dynamical simulations. The errors on the third column have been obtained from the variation of the numerical results on the maximal and the minimal interpolating function.}
\label{tab:mv/mrho_quenc}
\end{table}

\section{The chiral condensate anomalous dimension}
\label{sect:scaling}
The hyperscaling scenario supported by our data seems to imply the
existence of an infrared fixed point. However, since the evidence for
the hyperscaling and the locking of the mesonic and gluonic spectra
is still over a small range of $am$, simulations at
smaller masses and large volumes are needed to confirm the trend
identified so far.

If the theory is IR conformal, all the
spectral quantities scale as $m^{\rho}$ for a unique value of
$\rho = 1/(1+\gamma_*)$, with $\gamma_*$ the anomalous dimension of
the condensate. Hence, in this case $\gamma_*$ is physically
well-defined. 

In order to build phenomenologically viable Technicolor models, a large
anomalous dimension is generally required. From a purely theoretical
point of view 
$0 \le \gamma_* \le 2$, where $\gamma_*=0$ corresponds to the
non-interacting case and $\gamma_* = 2$ is the bound imposed by
unitarity; a value $\gamma_* \approx 1$ might reconcile Technicolor
with high-precision data for the Standard Model. The determination of
$\gamma_*$ is then one of the goals of lattice simulations of BSM
strong dynamics.

\begin{figure}[ht]
\begin{minipage}[b]{0.45\textwidth}
\centering
\includegraphics*[width=\textwidth]{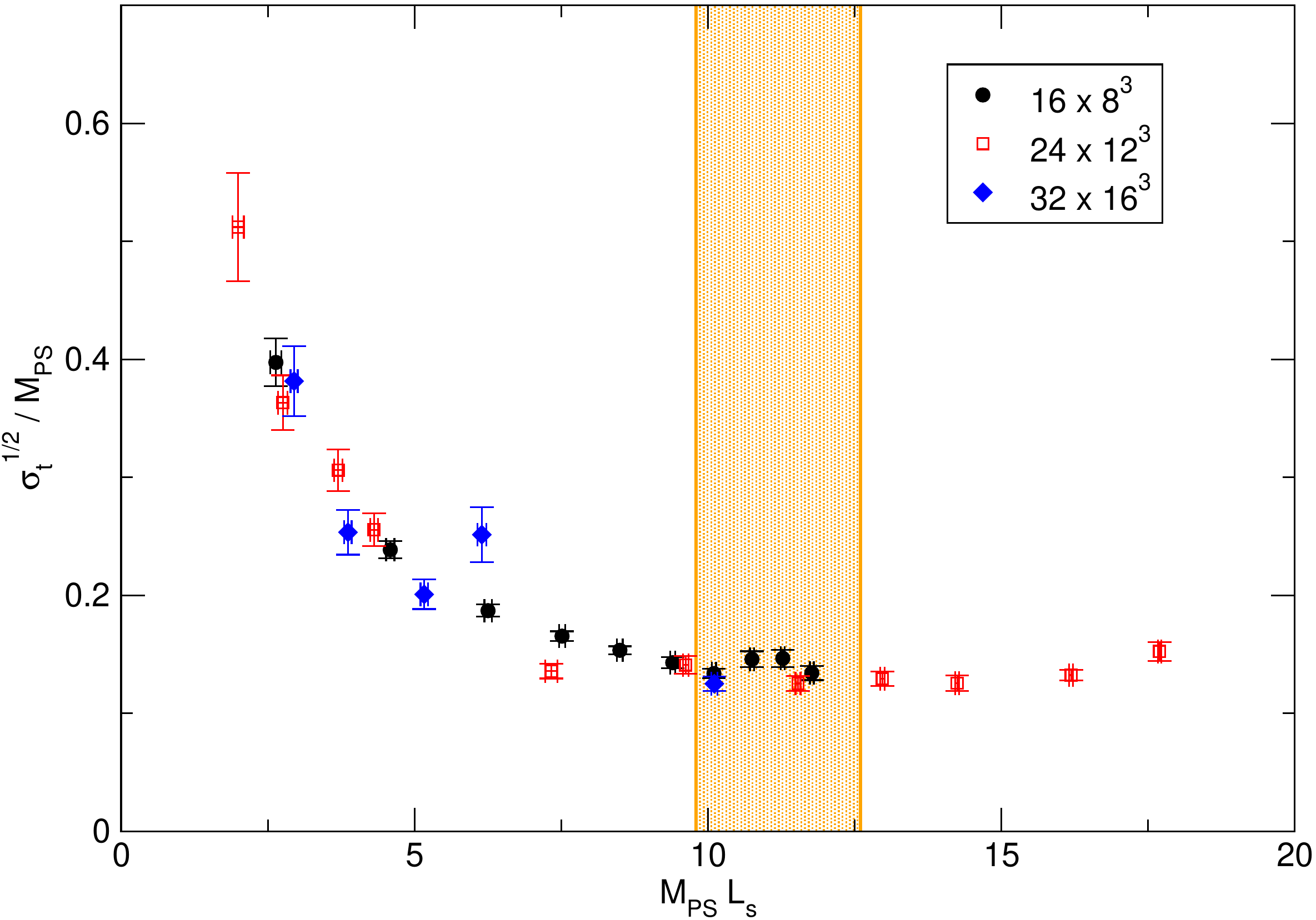}
\caption{The ratio $\sigma_t^{1/2}/M_\mathrm{PS}$ as a function of
  $M_\mathrm{PS} L_s$ at various lattice sizes. In the shaded region
  the crossover between the S and the A phase takes place.}
\label{fig:st}
\end{minipage}
\hspace{0.05\textwidth}
\begin{minipage}[b]{0.45\textwidth}
\centering
\includegraphics*[width=\textwidth]{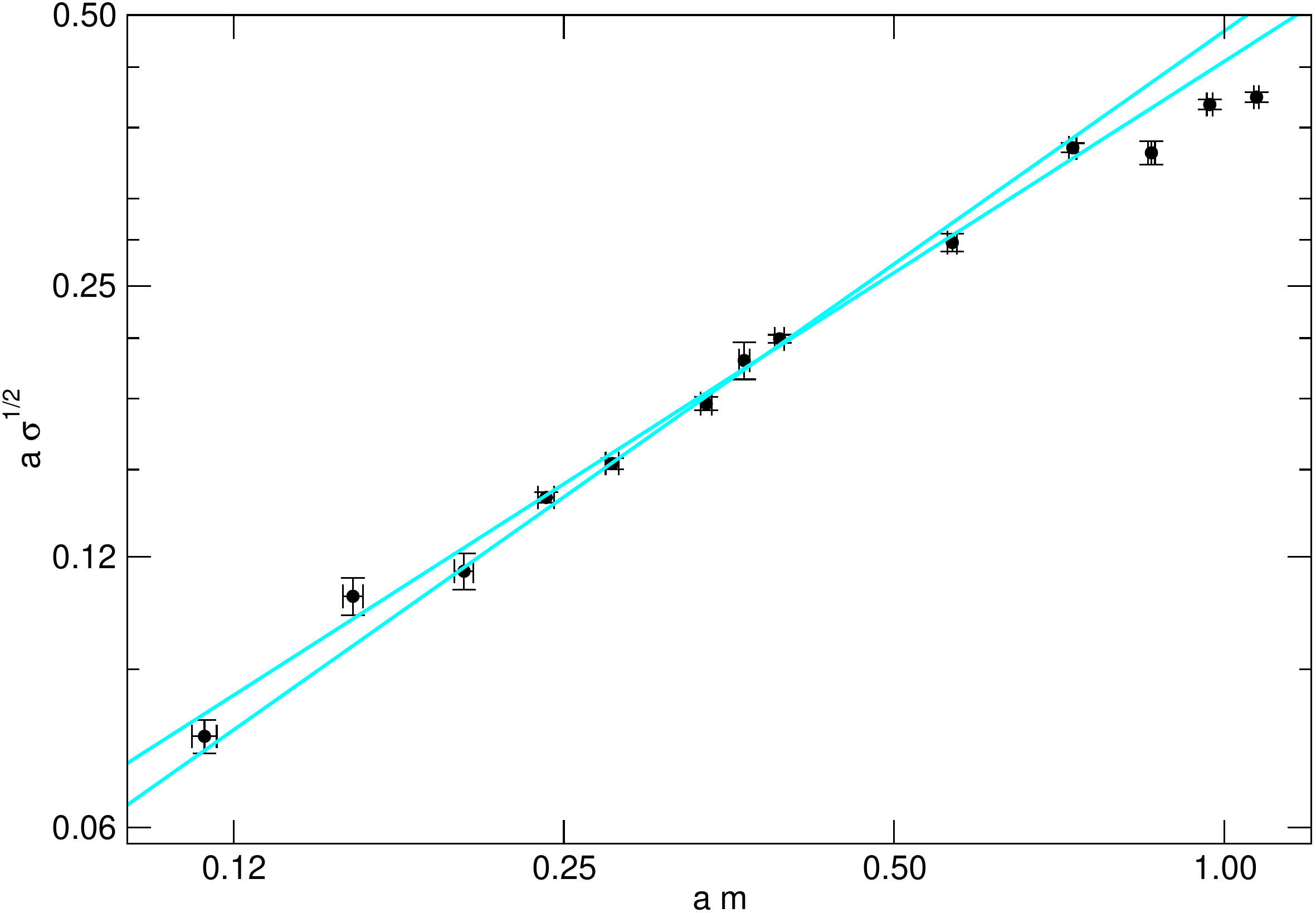}
\caption{$a\sigma^{1/2}$ as a function of $am$. A fit of the data to
  Eq.~\eqref{scaling} is also shown. In particular, the two lines
  represent the extremal values $\gamma_* = 0.16$ and $\gamma_* =
  0.28$.}
\label{fig:sqrtsigma_vs_mpcac}
\end{minipage}
\end{figure}

In order for us to be able to extract a scaling exponent, the
simulations must be performed in a region of sufficiently small
masses. The exact extent of the scaling region (which also depends on
the observable being analyzed) is only known {\em a posteriori}. On
the lattice, the problem is complicated by the explicit breaking of
conformal invariance due to the finite size of the system. This can
however be turned into a powerful tool for determining the exponent of
the scaling with the mass using a technique commonly known in
Statistical Mechanics as {\em Finite Size Scaling} (FSS). FSS states
that the dimension of the system is a relevant scaling variable with
mass dimension -1. Hence, the asymptotic scaling formula
\begin{eqnarray}
\label{scaling}
a M_X \propto (am)^{\rho} \ , \qquad \rho = 1/(1+\gamma_*) \ ,
\end{eqnarray}
where $M_X$ is a spectral quantity of the system, on a finite
lattice of spatial extension $L_s = a N_s$ and in the regime $L_s \to \infty$
and $m\to 0$ becomes
\begin{eqnarray}
M_X L_s =  f\left( x \right) \ , \qquad x = N_s (am)^{\rho} \ ,
\end{eqnarray}
i.e. the product $M_X L_s$ is a universal function of the scaling
variable $x$. A simple consequence is that the ratio of two spectral
quantities is expected to be a universal function of $M_X L_s$ for any
spectral quantity $M_X$. Note that this is true for both the S and the
A-phases of the system. In Fig.~\ref{fig:st} we show the ratio
$\sigma_t^{1/2}/M_\mathrm{PS}$ as a function of $M_\mathrm{PS} L_s$. The
universality of the ratio is verified up to values of $M_\mathrm{PS} L_s
\simeq 12$.

If the system is in the scaling region (for which we have support from
our data) and the infinite-volume estimates for spectral quantities are
correct, Eq.~\eqref{scaling} can be used for determining
$\gamma_*$. We have shown in Sect.~\ref{sect:lattice:string} that our
determination of the string tension is reasonably under
control. Hence, we perform a fit of our data for $\sigma^{1/2}$ using
Eq.~\eqref{scaling}. With a good quality of the fit (see
Fig.~\ref{fig:sqrtsigma_vs_mpcac}), we find $\gamma_* = 0.22(6)$. Both
horizontal and vertical data errors have been taken into account by
implementing a bootstrap procedure.  The fit has been performed on the
lightest four points, and then progressively increasing the fitting
region to include ten points (all but the last three in
Fig.~\ref{fig:sqrtsigma_vs_mpcac}). No systematic trend has been
observed when enlarging the fitting region. The quoted value of
$\gamma_*$ is a conservative estimate compatible with all the values
obtained using the fitting procedure described above.

The value we find for $\gamma_*$ is compatible with determinations
obtained in the same theory using other quantities (e.g. related to
mesonic physics~\cite{noi}) or independent techniques like the
Schr\"oedinger functional~\cite{Bursa:2009we}. These results clearly
favor the existence of a genuine IR fixed point for this theory.  The
fact that independent measurements of $\gamma_*$ fall all in the same
window of values is a clear message for model building.

\section{Conclusions}
\label{sect:conclusions}
In this work, using numerical simulations of the lattice model for
several sizes of the system and a wide range of fermion masses, we
have shown that at sufficiently low masses the spectrum of Minimal
Walking Technicolor is consistent with the existence of an infrared
fixed point.  In particular, for this specific realization of locking,
the theory at large distances is isospectral to a Yang-Mills SU(2)
theory, where the dynamically generated scale of the pure gauge theory
is determined by the fermion mass in MWT, and turns out to be smaller
than the latter.
To confirm this scenario would require to extend our study to much
smaller fermion masses, down to values that are not accessible at
present to lattice simulations. Another technical limitation of our
study is the simulation at a fixed value of $\beta$: in order to
verify that lattice artefacts are not distorting the physical picture,
further studies closer to the continuum limit should be performed.

Assuming the existence of the IR fixed point, we have determined the
anomalous dimension of the condensate, which is found to be $\gamma_*
= 0.22(6)$. This value is in agreement with other independent
determinations, which strengthen the conclusions that the theory is
infrared conformal. The value of $\gamma_*$ for this theory is
probably too small for conventional Technicolor scenarios, although
alternative scenarios compatible with a small anomalous dimension can
be devised (see e.g.~\cite{Evans:2005pu}). It would be desirable to
include larger lattices in our FSS analysis.

Finally, we notice that a FSS analysis performed in a SU(3) gauge
theory with two fermion flavors in the two-index symmetric
representation (which for SU(2) coincides with the adjoint
representation) finds $\gamma_* \simeq
0.5$~\cite{DeGrand:2009hu}. Assuming that, as stated in
Refs.~\cite{Shamir:2008pb,DeGrand:2008kx}, that this theory is
infrared conformal (however, see
Refs.~\cite{Fodor:2009ar,Kogut:2010cz} for alternative scenarios),
this might imply that $\gamma_*$ for two-index symmetric fermions is
an increasing function of the number of colors $N$. If this is the
case, it would be interesting to determine whether $\gamma_*$ becomes
of order one for large enough values of $N$ and whether $\gamma_*$
also increases with $N$ for adjoint fermions.

\begin{acknowledgments}
The numerical calculations presented in this work have been performed
on the BlueC supercomputer at Swansea university, on a Beowulf cluster
partly funded by the Royal Society and on the Horseshoe5 cluster
at the supercomputing facility at the University of Southern Denmark (SDU) 
funded by a grant of the Danish Centre for Scientific Computing for the project 
``Origin of Mass'' 2008/2009.
We thank C. Allton, J. Cardy, F. Knechtli, C. McNeile, M. Piai and F. Sannino for 
useful and fruitful discussions about various aspects related to this paper.
We thank the organizers and participants of the workshop ``Universe in a box'',
Lorentz Center, Leiden, NL, August 2009, where some results contained in this
paper were firstly presented and discussed.
A.P. thanks the groups at CERN, Columbia U., Maryland U.,
Colorado U., Washington U., LLNL, SLAC, Syracuse U. for warmily hosting him
and for useful and stimulating discussions about several aspects of this work.
Our work has been partially supported by STFC under contracts PP/E007228/1 and
ST/G000506/1. B.L. is supported by the Royal Society, A.P. is supported by STFC.
A.R. thanks the Deutsche Forschungsgemeinschaft for financial support.
\end{acknowledgments}

\bibliography{tgsu2}
\end{document}